\documentclass[]{aastex631}
\usepackage{amsmath}

\begin{document}

\title{21 cm Intensity Mapping with the DSA-2000}

\author[0000-0003-4980-2736]{Ruby Byrne}
\affiliation{Cahill Center for Astronomy and Astrophysics, California Institute of Technology, Pasadena CA 91125, USA}
\author{Nivedita Mahesh}
\affiliation{Cahill Center for Astronomy and Astrophysics, California Institute of Technology, Pasadena CA 91125, USA}
\author{Gregg W. Hallinan}
\affiliation{Cahill Center for Astronomy and Astrophysics, California Institute of Technology, Pasadena CA 91125, USA}
\author[0000-0002-7587-6352 ]{Liam Connor}
\affiliation{Cahill Center for Astronomy and Astrophysics, California Institute of Technology, Pasadena CA 91125, USA}
\author{Vikram Ravi}
\affiliation{Cahill Center for Astronomy and Astrophysics, California Institute of Technology, Pasadena CA 91125, USA}
\author{T. Joseph W. Lazio}
\affiliation{Jet Propulsion Laboratory, California Institute of Technology, 4800 Oak Grove Drive, Pasadena, CA 91109, USA}



\begin{abstract}

Line intensity mapping is a promising probe of the universe's large-scale structure. We explore the sensitivity of the DSA-2000, a forthcoming array consisting of over 2000 dishes, to the statistical power spectrum of neutral hydrogen's 21 cm emission line. These measurements would reveal the distribution of neutral hydrogen throughout the near-redshift universe without necessitating resolving individual sources. The success of these measurements relies on the instrument's sensitivity and resilience to systematics. We show that the DSA-2000 will have the sensitivity needed to detect the 21 cm power spectrum at $z \approx 0.5$ and across power spectrum modes of $0.03-35.12 \, h/\text{Mpc}$ with $0.1 \, h/\text{Mpc}$ resolution. We find that supplementing the nominal array design with a dense core of 200 antennas will expand its sensitivity at low power spectrum modes and enable measurement of Baryon Acoustic Oscillations (BAOs). Finally, we present a qualitative discussion of the DSA-2000's unique resilience to sources of systematic error that can preclude 21 cm intensity mapping.

\end{abstract}



\section{Introduction} \label{sec:intro}

Line intensity mapping is an emerging approach for characterizing large volumes of the universe \citep{Battye2004, Chang2008, Wyithe2008, Kovetz2017}. It involves measuring the statistical properties of emission across different angular scales, enabling surveys of large swaths of sky with relatively low-resolution instruments. While line intensity mapping experiments aim to measure a variety of emission lines---including of CO, CII, and Ly$\alpha$---of particular interest is the narrow hyperfine transition line of hydrogen, called ``21 cm emission'' due to its rest-frame wavelength of 21 cm.

21 cm emission is the principal tracer of neutral hydrogen gas (HI) throughout the universe, and measurement of the signal across redshift would constrain cosmological evolution and galactic dynamics. As the signal is quite faint and contaminated by bright foreground emission, it is most accessible at low redshifts. In an era of large galaxy surveys---such as the Sloan Digital Sky Survey (SDSS; \citealt{York2000, Blanton2017}), the extended Baryon Oscillation Spectroscopic Survey (eBOSS; \citealt{Dawson2016}), the Dark Energy Spectroscopic Instrument (DESI; \citealt{DESI2016, Moon2023}), Euclid \citep{Euclid2022}, and the Legacy Survey of Space and Time (LSST) from the Rubin Observatory \citep{Ivezic2019}---21 cm intensity mapping has the advantage that it is not biased toward massive galaxies and instead probes the density field contribution from faint, unresolved galaxies. Additionally, mapping the narrow 21 cm emission line provides excellent redshift resolution beyond that achievable with galaxy surveys \citep{Bull2015, Santos2015, Villaescusa-Navarro2018, Chen2021}. 

In recent years, the near-redshift 21 cm signal has been measured in cross-correlation with galaxy surveys with GBT \citep{Chang2010,Masui2013,Wolz2021}, Parkes \citep{Anderson2018}, CHIME \citep{CHIME2022}, and MeerKAT \citep{Cunnington2022}; \citet{Chime2023} detected the 21 cm signal in correlation with the Lyman-$\alpha$ forest. \citet{Paul2023} reported the first-ever detection of the 21 cm autocorrelation power spectrum with the MeerKAT telescope. The detection corresponds to a redshift range of $0.32 \le z \le 0.44$ and power spectrum modes of $0.48 \, h/\text{Mpc} \le k \le 11.21 \, h/\text{Mpc}$.

This paper explores measurement of the 21 cm power spectrum with the DSA-2000, a forthcoming 2000-element radio array to be built in Nevada \citep{Hallinan2019}. The DSA-2000 will operate at frequencies of 0.7–2 GHz and is optimized for fast survey speeds. It will leverage the novel Radio Camera imaging pipeline for real-time imaging, allowing for efficient processing of fully cross-correlated interferometric data. As a multi-purpose observatory, the DSA-2000 has wide-reaching science applications including fast radio burst (FRB) detection and localization, weak lensing studies \citep{Connor2022}, multi-messenger astronomy, pulsar timing with NANOGrav \citep{Wahl2022}, and more. We show that it will be a powerful instrument for near-redshift 21 cm intensity mapping.

Intensity mapping experiments require both exquisite sensitivity and excellent systematics control. Sensitivity is crucial for detecting the faint signal, and in this paper we explore the expected sensitivity of the DSA-2000 to the 21 cm power spectrum signal. We show that, due to its many antennas and fast survey speeds, the DSA-2000 will have the necessary sensitivity to measure the 21 cm power spectrum across a large range of power spectrum modes with just a few minutes of data. In a forthcoming paper, we predict the resulting constraints on cosmological parameters (Mahesh et al., in prep.).

However, sensitivity alone is not sufficient for intensity mapping, as systematics can quickly overwhelm the signal and prevent separation of the astrophysical foregrounds. For this reason, instrument characterization through calibration is crucial. While several experiments in the field have the needed sensitivity to in principle measure the 21 cm power spectrum, the DSA-2000 is a uniquely calibratable instrument, owing once again to its many antennas and its pseudo-random array layout. Furthermore, its imaging capabilities will enable excellent foreground characterization and separation with minimal spectral leakage. This will make the DSA-2000 a powerful addition to the field of low-redshift 21 cm intensity mapping.

\section{The DSA-2000}

The DSA-2000 is a survey instrument that builds upon its pathfinders, the DSA-10 and DSA-110 \citep{Kocz2019, Ravi2022}. Over the course of 4-month observing seasons, the DSA-2000 will survey the entire sky above a declination of -30$^\circ$, delivering images at a spatial resolution of 3.5 arcseconds and with a sensitivity of 2 $\mu$Jy/beam \citep{Hallinan2019}. It will operate as a fast survey complement to the Square Kilometre Array mid (SKA-mid) and the Next Generation Very Large Array (ngVLA), with approximately ten times the survey speed of each of those forthcoming instruments.

\begin{figure}
    \centering
    \includegraphics[width=\columnwidth]{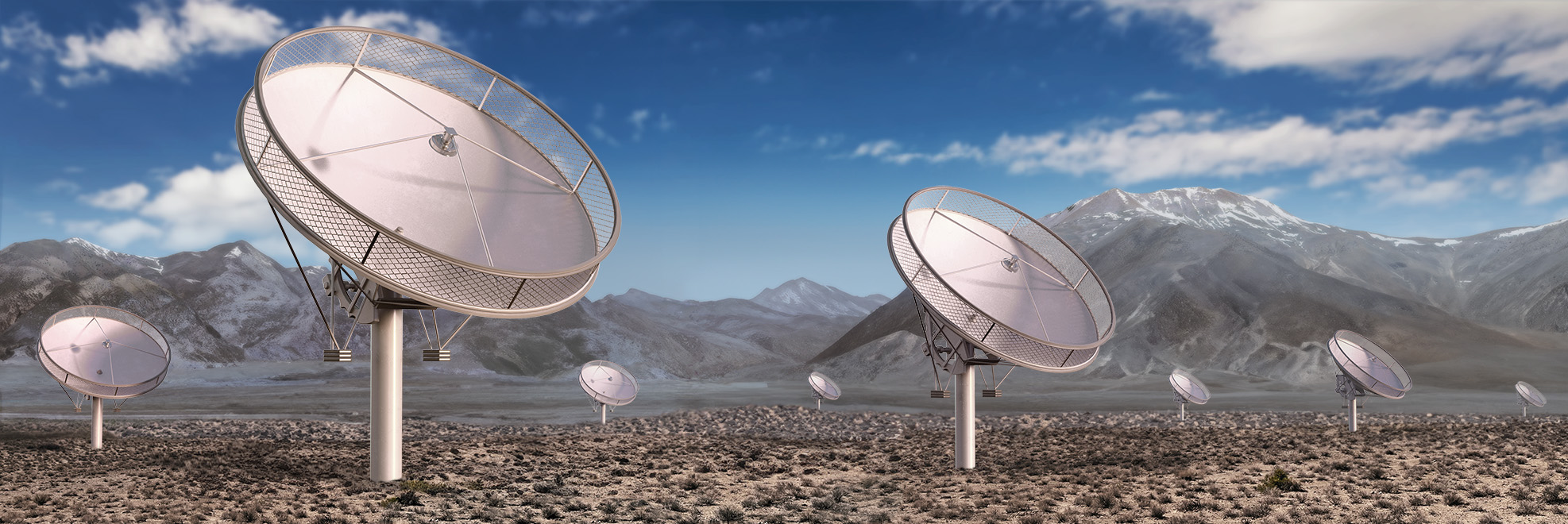}
    \caption{Artist's rendering of the DSA-2000 antennas. Each dish has a diameter of 5 m, and the antennas are fully steerable in azimuth and elevation. The array will consist of approximately 2000 such antennas, configured as pictured in Figure \ref{fig:antlocs}.}
    \label{fig:dish_image}
\end{figure}

The DSA-2000 achieves this combination of resolution, sensitivity, and survey speed with an unprecedented 2000 antenna dishes, pictured in Figure \ref{fig:dish_image}. This is enabled by two key technological advances. First, innovative ambient-temperature low-noise amplifiers allow for inexpensive dish construction \citep{Weinreb2021}. Next, advances in graphics processing unit (GPU) technology in recent years will enable real-time imaging of the array's over 2 million visibilities through a data processing pipeline dubbed the Radio Camera.

Each of the DSA-2000 antennas will be constructed from a 5 m hydroformed aluminum dish. The antennas will be fully steerable in both azimuth and elevation, differing from the pathfinder DSA-10 and DSA-110 dishes that are steerable in elevation only. A quad-ridge horn feed at the focus of each dish will allow for wideband dual-polarization measurements and will house the low-noise amplifier described in \citealt{Weinreb2021}. Signals will be transported via RF-over-fiber to a central processing building, where they will be digitized, correlated, and imaged.

\begin{figure}
    \centering
    \includegraphics[width=0.8\columnwidth]{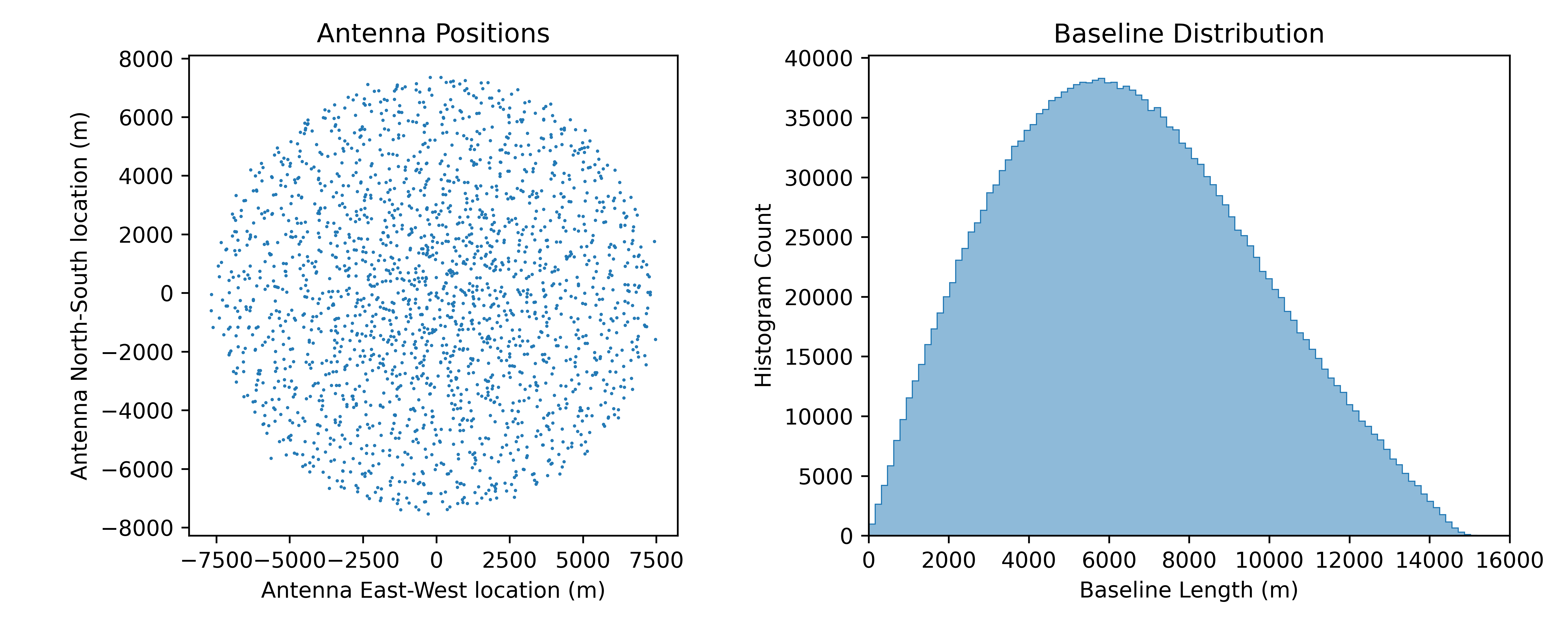}
    \caption{Plot of the proposed DSA-2000 antenna locations (left) and the resulting distribution of baseline lengths (right). The DSA-2000 array layout is designed to maximize \textit{uv} sampling and produce a well-behaved PSF, plotted in Figure \ref{fig:psf}.}
    \label{fig:antlocs}
\end{figure}

\begin{figure}
    \centering
    \includegraphics[width=0.5\columnwidth]{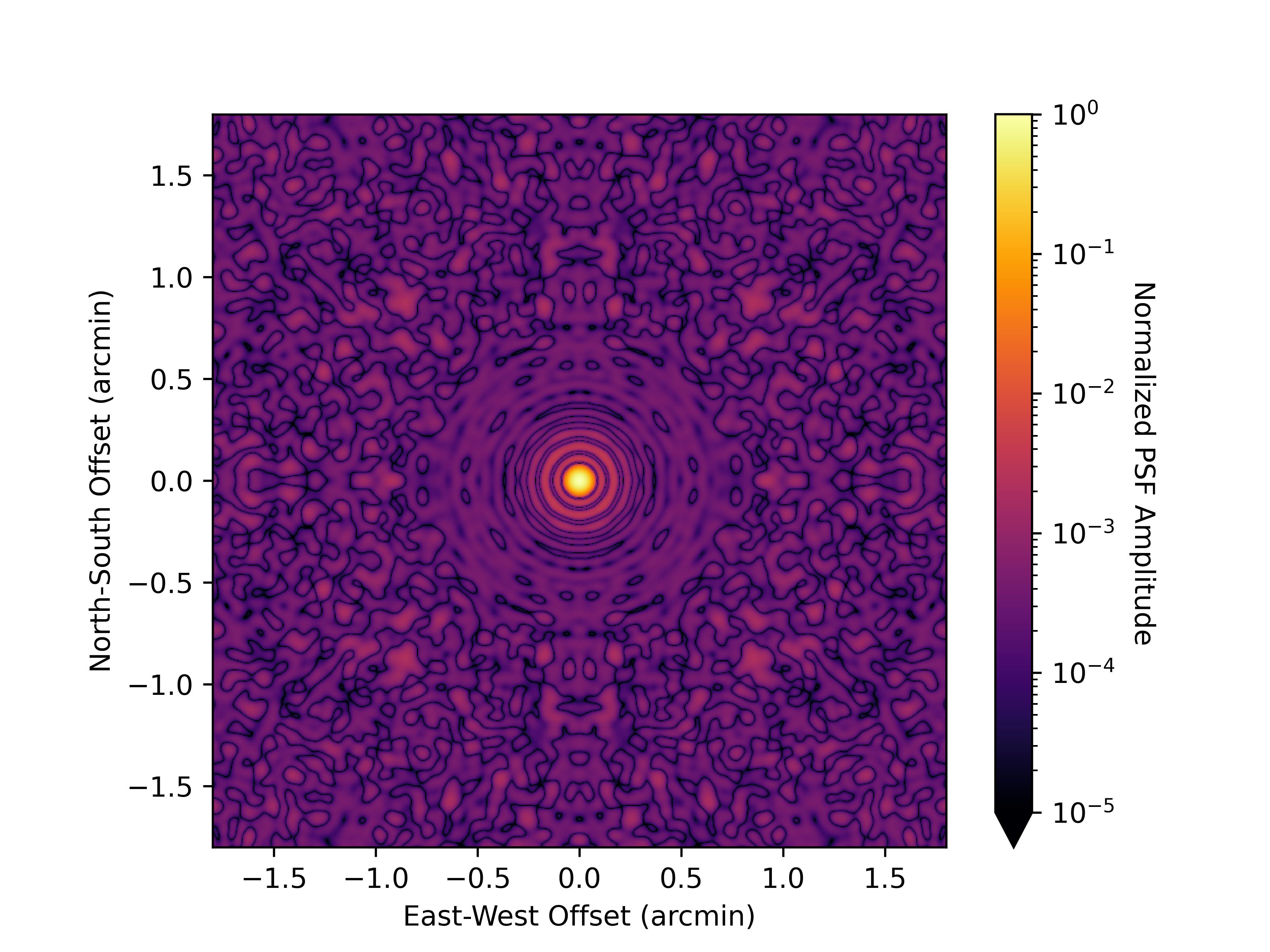}
    \caption{Plot of the normalized amplitude of the PSF of the DSA-2000, based on the proposed antenna positions plotted in Figure \ref{fig:antlocs}. The DSA-2000 is designed to have a highly peaked PSF with low sidelobe amplitudes. Here we have normalized the PSF such that the peak value is 1. We plot the PSF of a snapshot, zenith-pointed observation at a single frequency of 1.06 GHz. Frequency averaging and synthesis rotation will further suppress the PSF sidelobes.}
    \label{fig:psf}
\end{figure}

The Radio Camera data processing pipeline relies on the DSA-2000's exquisite imaging fidelity, owing to its many antennas and well-behaved point spread function (PSF). The pseudo-random array configuration, pictured in Figure \ref{fig:antlocs}, is designed to produce a PSF with very low amplitude sidelobes, plotted in Figure \ref{fig:psf}. This obviates the need for computationally costly visibility-based deconvolution \citep{Connor2022}. It also means that the DSA-2000 will have excellent \textit{uv} coverage for power spectrum analyses such as 21 cm intensity mapping and that it will exhibit minimal spectral structure, reducing a dominant systematic for 21 cm analyses.

\section{Defining the Predicted Signal}
\label{s:predicted_ps}

\begin{table}
\centering
\begin{tabular}{|c | c | c|} 
 \hline
 \textbf{Variable} & \textbf{Description} & \textbf{Value} \\ 
 \hline
 \hline
 $h$ & Dimensionless Hubble constant & 0.6766 \\ 
 \hline
 $H_0$ & Hubble constant & $h \times 100$ km s$^{-1}$ Mpc$^{-1}$\\ 
 \hline
 $\Omega_m$ & Matter density & 0.3111 \\ 
 \hline
 $\Omega_k$ & Curvature constant & 0 \\ 
 \hline
 $\Omega_{\Lambda}$ & Dark energy density & 0.6889 \\
 \hline
 $\Omega_B$ & Baryon density & 0.0490 \\
 \hline
 $b$ & HI halo bias & 0.75 \\
 \hline
 $f_\text{HI}$ & HI mass fraction & 0.015 \\
 \hline
\end{tabular}
\caption{
Cosmological parameter values used throughout this paper. Estimating the 21 cm signal requires fiducial values for the cosmological constants, listed here. We assume flat cosmology ($\Omega_k=0$) and use $H_0$, $\Omega_m$, $\Omega_\Lambda$, and $\Omega_B$ from the Planck 2018 results \citep{Planck2020}. $b$ and $f_\text{HI}$ are from \citealt{Pober2013a}.
}
\label{tab:cosmological_parameters}
\end{table}

While the DSA-2000 has a myriad of scientific applications, this paper concerns detection of the power spectrum of 21 cm emission. In this section, we define the predicted signal at the redshifts of interest.

The 21 cm power spectrum at low redshift is given by
\begin{equation}
    P_{21}(k, z) = [T_{21}(z)]^2 b^2 P_\text{mat}(k,z)
\label{eq:theory_ps}
\end{equation}
where $T_{21}(z)$ is the mean 21 cm brightness temperature at redshift $z$, $P_\text{mat}(k,z)$ is the matter power spectrum, and $b$ is the bias factor for HI halos. We represent $T_{21}(z)$ as
\begin{equation}
    T_{21}(z) \approx 0.084\text{mK}\frac{(1+z)^2 h}{\sqrt{\Omega_m(1+z)^3+\Omega_\Lambda}} \frac{\Omega_B}{0.044} \frac{f_\text{HI}(z)}{0.01}
\label{eq:brightness_temp}
\end{equation}
\citep{Pober2013a}. Here $f_\text{HI}(z)$ is the mass fraction of HI with respect to overall cosmological baryon content. Table \ref{tab:cosmological_parameters} presents the cosmological parameter values used throughout this paper.

We calculate the matter power spectrum $P_\text{mat}(k, z)$ at redshift $z=0.5$ and $k= 0-11.8 \, h/\text{Mpc}$ with the Code for Anisotropies in the Microwave Background (CAMB; \citealt{Lewis2000}).\footnote{\texttt{https://lambda.gsfc.nasa.gov/toolbox/camb\_online.html}} We define the normalized theoretical power spectrum $P_\text{theory}(k, z)$, with units mK$^2$, as
\begin{equation}
    P_\text{theory}(k, z) = \frac{k^3}{2 \pi^2} P_{21}(k, z).
\label{eq:theory}
\end{equation}
Figures \ref{fig:therm_noise}, \ref{fig:sample_stddev}, and \ref{fig:error_bars} plot $P_\text{theory}(k, z)$ in black. The dashed black line indicates extrapolated values at high $k$.

\section{Estimating the Measurement Sensitivity}
\label{s:sensitivity}

In this section, we quantify the sensitivity of the DSA-2000 by comparing the predicted statistical uncertainty of its measurement to the expected 21 cm power spectrum defined above in \S\ref{s:predicted_ps}. This analysis determines the measurement capabilities of the DSA-2000 in the absence of systematics. Adequate sensitivity is a necessary but not sufficient condition for achieving the measurement, and \S\ref{s:systematics} discusses anticipated sources of systematic error.

Our analysis is based on a delay spectrum approach discussed in, for example, \citealt{Parsons2012} and \citealt{Kolopanis2019} and described in detail below in \S\ref{s:delay_spectrum}. This approach offers a simple, analytic estimate of the impact of thermal noise on the 21 cm power spectrum.

We consider three noise sources: thermal noise, sample variance, and shot noise. The first benefits from a low system temperature, $T_\text{sys}$, and can be reduced through long time integrations, whereas the sample variance and shot noise are mitigated by measuring large swaths of the sky.

Our analysis is simplified by omitting the effects of synthesis rotation. Synthesis rotation exploits the Earth's rotation to improve the instrument's \textit{uv} sampling. The degree of synthesis rotation depends on the survey strategy used, and we neglect it here to provide a more general estimate of the instrument's performance. Furthermore, our analysis omits consideration of techniques commonly used in the field to reduce noise bias, such as correlating interleaved time steps or removing baseline autocorrelations (see \citet{Morales2023} for an overview of those techniques). We use the cylindrically averaged 1D power spectrum as our figure of merit, and we therefore do not consider non-isotropic effects, such as the ``fingers of God,'' that break symmetry between modes parallel and perpendicular to the line of sight.

We consider the effect of foreground contamination and use an aggressive foreground avoidance strategy that masks power spectrum modes likely to be affected by foreground emission. However, in \S\ref{s:discussion} we note that the DSA-2000 may be able to use a less stringent foreground avoidance approach that recovers some of these masked modes.

The sensitivity of a power spectrum analysis depends on data volume that contributes to each measurement bin. Our result therefore scales with the chosen redshift and power spectrum bin sizes. Larger bins produce more sensitive measurements but sacrifice measurement resolution. In this section, we use linearly-spaced power spectrum bins of widths of either $\Delta k = 0.1 \, h/\text{Mpc}$ (Figures \ref{fig:therm_noise}, \ref{fig:sample_stddev}, and \ref{fig:error_bars}) or $\Delta k = 0.03 \, h/\text{Mpc}$ (Figures \ref{fig:bao_error_bars}, \ref{fig:bao_error_bars_offaxis}, and \ref{fig:bao_error_bars_core}). We use the full frequency band of $0.7-1.43$ GHz, corresponding to a redshift bin of $0 \le z \le 1$.

All software underlying this analysis is available open-source on \textit{GitHub}\footnote{See \texttt{https://github.com/rlbyrne/PSsensitivity}. The software is available open-source under a BSD 2-Clause ``Simplified'' License and archived with Zenodo \citep{pssensitivity2024}.} \citep{pssensitivity2024}.

\subsection{The Delay Spectrum Analysis}
\label{s:delay_spectrum}

Here we detail the data analysis pipeline that underlies our thermal sensitivity analysis, presented below in \S\ref{s:thermal_var}. 

Intensity mapping analyses broadly fall into two categories: delay spectrum analyses and imaging power spectrum analyses (for a detailed comparison of these methods, see \citealt{Morales2019} or \citealt{Liu2020}). An imaging power spectrum analysis requires a reconstructed \textit{uv} plane, created by gridding visibilities and analogous to the angular Fourier transform of the reconstructed sky image. In constrast, the delay spectrum analysis bypasses gridding. Instead, the visibilities are Fourier transformed across frequency, squared, and then binned to produce a power spectrum estimate. This approach has the benefit that it is simple, computationally efficient, and enables analytic error propagation, and this paper therefore uses a delay spectrum analysis approach to derive the measurement sensitivity. In practice, however, we expect intensity mapping with the DSA-2000 to use an imaging power spectrum approach, as it better synergizes with the Radio Camera imaging pipeline and does not require storing the raw visibilities. While the noise properties of the imaging and delay spectrum approaches are not identical, we expect that our delay spectrum-based sensitivity estimate is a reasonable approximation of the noise for any intensity mapping analysis. 

This delay spectrum analysis consists of three steps: Fourier transforming the visibilities (\S\ref{s:ft}), implementing foreground avoidance to remove the contaminating foreground signal (\S\ref{s:foreground_masking}), and finally combining the visiblities to form the power spectrum estimate (\S\ref{s:binning}). In \S\ref{s:thermal_var} we describe how we propagate the expected thermal noise through this pipeline.

\subsubsection{Fourier Transform}
\label{s:ft}

We represent the visibility formed by correlating antennas $i$ and $j$ as $v_{ij}(f)$, where $f$ denotes frequency. The first step in the delay spectrum analysis involves performing a discrete Fourier transform across frequency for each visibility, which we define as follows:
\begin{equation}
    \tilde{v}_{ij}(\eta) = \sum_f W(f) v_{ij}(f) e^{-2\pi i \eta f}
\label{eq:ft}
\end{equation}
Here $\eta$ represents delay, the Fourier dual of frequency with units of time, and the tilde ($\,\tilde{}\,$) denotes the Fourier transformed quantity. $W(f)$ is the anti-aliasing window function (e.g., Blackman-Harris or Tukey). To preserve power, the window function is normalized such that
\begin{equation}
    \sum_f W^2(f) = N_\text{freq},
\label{eq:window_function_normalization}
\end{equation}
where $N_\text{freq}$ is the number of frequency channels.

In this analysis we neglect the effects of flagging and assume evenly-spaced frequency channels, such that a simple discrete Fourier transform suffices. The introduction of frequency-dependent data excision (``flagging’'), for mitigation of radio-frequency interference (RFI) and other data contaminants, could result in deleterious spectral mode-mixing \citep{Wilensky2021}, if not handled correctly. At minimum, calculating the delay spectrum following data excision requires an approach that accounts for gaps in the spectral data, such as the Lomb-Scargle periodogram (see \citealt{Lomb1976}, \citealt{Scargle1982}, and \citealt{VanderPlas2018} for a description of the technique; \citealt{Jacobs2016} describes its application to 21 cm cosmology with the Murchison Widefield Array). Beyond that, other algorithms may result in better spectral performance with reduced mode-mixing. These include iterative deconvolution techniques such as the CLEAN algorithm \citep{Hogbom1974, Parsons2014}; Markov Chain Monte Carlo (MCMC) algorithms such as Gibbs sampling \citep{CHIME2022, Kennedy2023}; and frequency inpainting with Gaussian Process Regression (kriging) \citep{Mertens2018, Trott2020}.

\subsubsection{Foreground Avoidance}
\label{s:foreground_masking}

Foreground mitigation strategies for 21 cm analyses fall into two categories: foreground subtraction and foreground avoidance. Foreground subtraction entails modeling and subtracting bright foreground sources, while foreground avoidance removes foreground power by masking the contaminated power spectrum modes. The latter approach works because the foregrounds are inherently spectrally smooth, owing to their synchrotron and free-free emission mechanisms, and therefore inhabit a compact region in power spectrum space. While many analyses in the field employ some combination of foreground subtraction and avoidance, it is infeasible to model the foreground emission with the fidelity necessary to rely on foreground subtraction alone. We therefore adopt a foreground avoidance approach in this analysis.

Spectrally smooth foregrounds inhabit a wedge-shaped region of 2-D power spectrum space, dubbed the ``foreground wedge'' and discussed extensively in the literature \citep{Morales2012, Trott2012, Vedantham2012, Pober2013b, Thyagarajan2013, Hazelton2013, Liu2014a, Liu2014b, Dillon2015, Morales2019}. The extent of the foreground wedge depends on the largest angle at which the instrument detects off-axis foreground emission. This angle is set nominally by the instrument's field of view. However, as instruments may have significant nonzero sensitivity beyond the field of view extent, many analyses in the field take a conservative approach to foreground avoidance and omit wedge modes down to the horizon.

We represent the foreground mask with a binary function $M_{ij}(\eta)$ where values of 0 correspond to masked modes. We define this function as
\begin{equation}
    M_{ij}(\eta) = \begin{cases}
    0, & |\eta| - \Delta \eta/2 \le (\sin \omega) \, |\boldsymbol{b}_{ij}|/c  \\
    1, & |\eta| - \Delta \eta/2 > (\sin \omega) \, |\boldsymbol{b}_{ij}|/c
    \end{cases}.
\end{equation}
Here $|\boldsymbol{b}_{ij}|$ is the length of the baselines formed by antennas $i$ and $j$ in units of meters and $c$ represents the speed of light. $\sin \omega$ sets the extent of the wedge, where $\omega$ corresponds to the maximum angle from the pointing center where instrument is sensitive to foregrounds. Setting $\omega = 90^\circ$ corresponds to a horizon cut. We assume the delay axis is discretized into evenly-spaced bins of width $\Delta \eta$, and $\eta$ corresponds to the value at the center of the bin. The term $\Delta \eta/2$ ensures that any delay bin containing foreground contamination will be masked.

\subsubsection{Binning and Averaging}
\label{s:binning}

The next step in the analysis pipeline is combining the Fourier transformed visibilities into power spectrum bins and averaging their squared magnitudes. Because the DSA-2000 is a non-regular array, visiblities cannot be coherently averaged before squaring. The resulting reconstructed power spectrum is given by
\begin{equation}
    \hat{P}(k) = \frac{\sum_{ij\eta \in k} M_{ij}(\eta) |\tilde{v}_{ij}(\eta)|^2}{\sum_{ij\eta \in k} M_{ij}(\eta)},
\label{eq:power_spectrum_est}
\end{equation}
where the $\hat{}$ symbol indicates the empirically estimated value. Here the sum is taken over all values that contribute to power spectrum bin $k$. $M_{ij}(\eta)$ represents the binary masking function described above in \S\ref{s:foreground_masking}, so the denominator of this expression simply counts the number of contributing values.

To determine which values contribute to each bin, we must convert the baseline length and delay $\eta$ into cosmological units. See Appendix \ref{app:cosmological_units} for a description of that conversion and Table \ref{tab:cosmological_parameters} for values of cosmological parameters used. In terms of the conversion factors $C_\parallel$ and $C_\perp$ defined in Appendix A, a visibility mode $\tilde{v}_{ij}(\eta)$ contributes to power spectrum bin $k$ if 
\begin{equation}
    k - \Delta k /2 < \sqrt{ \left( C_\perp |\boldsymbol{b}_{ij}| f / c \right)^2 + C_\parallel^2 \eta^2} < k + \Delta k /2.
\end{equation}
Here $\Delta k$ is the bin width, $c$ is the speed of light, $|\boldsymbol{b}_{ij}|$ is the baseline length, and $f$ is the frequency, which we define as the central frequency of the band.

\subsection{Thermal Noise Propagation}
\label{s:thermal_var}

\begin{table}
\centering
\begin{tabular}{| c | c | c|} 
 \hline
 \textbf{Variable} & \textbf{Description} & \textbf{Value} \\ 
 \hline
 \hline
 $T_\text{sys}$ & System temperature & 25 K \\ [0.5ex] 
 \hline
 $D_a$ & Antenna diameter & 5 m \\ 
 \hline
 $N_\text{ants}$ & Number of antennas & 2,048 \\ 
 \hline
 $N_\text{bls}$ & Number of baselines & 2,098,176 \\ 
 \hline
 $\Delta f$ & Frequency channel width & 130.2 kHz \\ 
 \hline
 $e_a$ & Aperture efficiency & 0.7 \\ 
 \hline
 $f_\text{min}$ & Minimum frequency & 0.7 GHz \\ 
 \hline
 $f_\text{max}$ & Maximum frequency & 1.43 GHz \\ 
 \hline
 FoV & Field of view at 1.06 GHz & 30.0 deg$^2$ \\ 
 \hline
 $\Delta \Theta$ & Field of view diameter at 1.06 GHz & 6.18$^\circ$ \\ 
 \hline
\end{tabular}
\caption{Instrument parameter values used throughout this paper. While the DSA-2000 measures frequencies up to 2 GHz, the maximum frequency used in this paper refers to the 21 cm rest-frame frequency. Here we define the limits of the FoV at the 5\% beam level, where we have modeled the antennas as circular apertures with Airy disk beams.}
\label{tab:instrument_parameters}
\end{table}

We can now propagate the expected thermal noise through the analysis steps outlined above in \S\ref{s:delay_spectrum}.

The thermal noise of a visibility is quantified by the root-mean-square (RMS):
\begin{equation}
\operatorname{RMS}[v_{ij}(f)] = \frac{ \lambda^2 T_\text{sys}}{A_e \sqrt{\Delta f \tau}}
\end{equation}
where $\lambda$ is the observed wavelength, $T_\text{sys}$ is the system temperature, $\Delta f$ is the channel width, $\tau$ is the integration time, and $A_e$ is the effective collecting area \citep{Morales2010}. We approximate the effective collecting area as
\begin{equation}
    A_e = \frac{\pi}{4} D_a^2 e_a,
\end{equation}
where $D_a$ is the antenna diameter and $e_a$ is the aperture efficiency. A table of the antenna parameter values used in this paper is provided in Table \ref{tab:instrument_parameters}.

If we assume the thermal noise on each visibility is circularly Gaussian distributed, then we can define the thermal noise variance of each the real and imaginary components of the visibilities (denoted $\Re[v_{ij}(f)]$ and $\Im[v_{ij}(f)]$, respectively) as
\begin{equation}
    \sigma_\text{vis}^2 = \operatorname{Var}_\text{therm}\left( \Re[v_{ij}(f)]\right) = \operatorname{Var}_\text{therm}\left( \Im[v_{ij}(f)]\right) = \frac{1}{2} \left( \operatorname{RMS}[v_{ij}(f)] \right)^2.
\end{equation}
Here $\operatorname{Var}_\text{therm}$ denotes the thermal noise variance. If we further assume that the noise is independent on each frequency channel and propagate the thermal noise variance through the Fourier transform operation in Equation \ref{eq:ft}, we get that
\begin{equation}
    \operatorname{Var}_\text{therm}\left( \Re[\tilde{v}_{ij}(\eta)]\right) = \operatorname{Var}_\text{therm}\left( \Im[\tilde{v}_{ij}(\eta)]\right) = \sigma_\text{vis}^2 \sum_f W^2(f).
\end{equation}
From the normalization of the window function in Equation \ref{eq:window_function_normalization}, this reduces to
\begin{equation}
    \operatorname{Var}_\text{therm}\left( \Re[\tilde{v}_{ij}(\eta)]\right) = \operatorname{Var}_\text{therm}\left( \Im[\tilde{v}_{ij}(\eta)]\right) = N_\text{freq} \sigma^2_\text{vis}.
\end{equation}
We thereby find that the noise on the Fourier transformed visibilities is independent of the window function used.

\begin{figure}
    \centering
    \includegraphics[width=0.6\columnwidth]{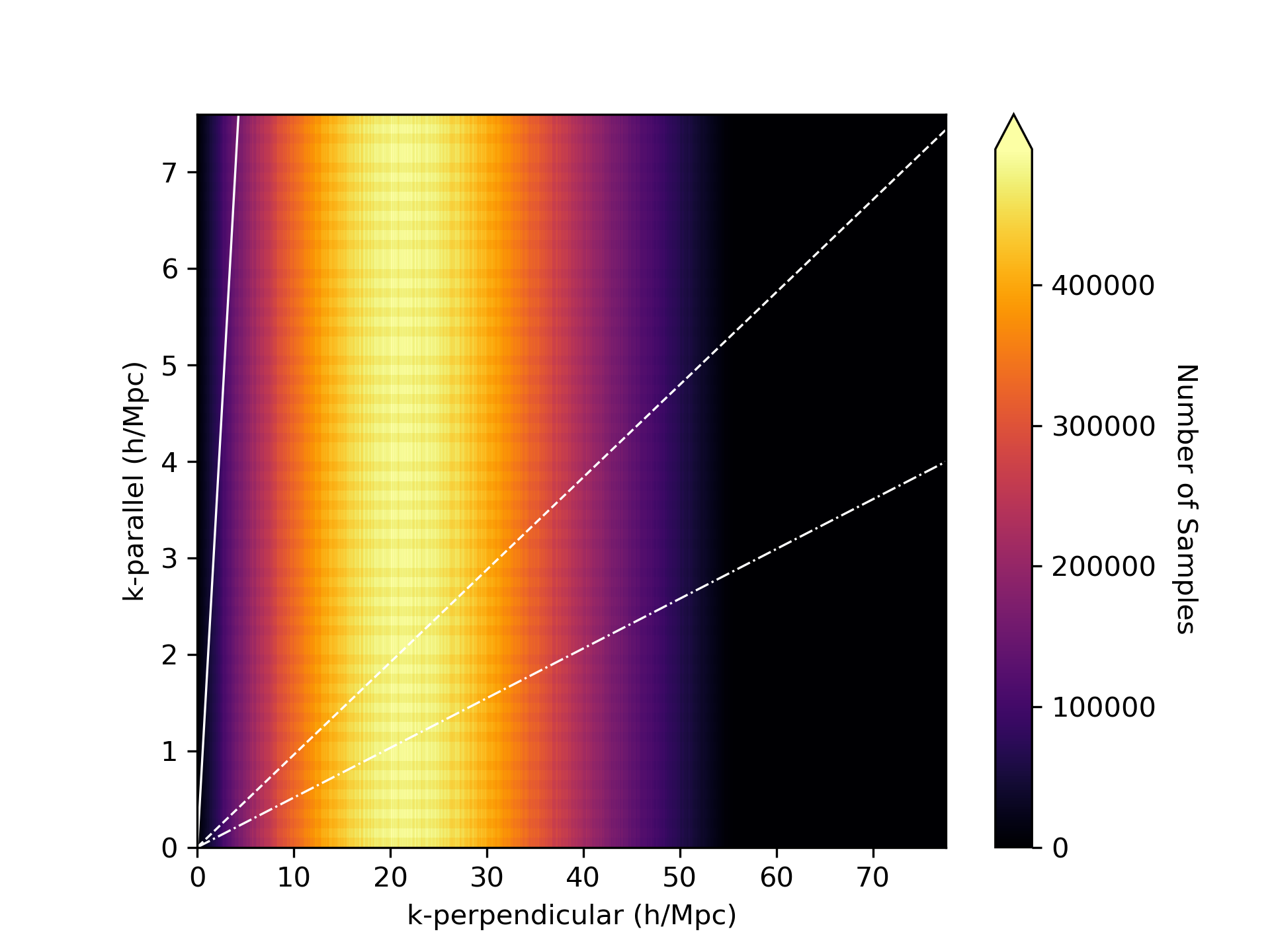}
    \caption{Plot of the sampling of 2D $(k_\perp, k_\parallel)$ space. The colorbar indicates the number of visibility measurements in each 0.1 $h$/Mpc $\times$ 0.1 $h$/Mpc bin. The $k_\perp$-dependence traces the baseline distribution plotted in Figure \ref{fig:antlocs}. The solid white line indicates the horizon extent of the foreground wedge, the dashed white line indicates the field of view extent of the foreground wedge (here we define the field of view as the region within the 5\% beam power level), and the dash-dotted line indicates the extent of the wedge within the beam half maximum. Using the horizon for foreground wedge exclusion eliminates most measurements and may be more conservative than necessary for effective foreground avoidance.}
    \label{fig:2d_samples}
\end{figure}

The Fourier transform is a linear operation, so it preserves the circular Gaussian noise distribution profile. It follows that the squared quantity has noise variance
\begin{equation}
    \operatorname{Var}_\text{therm}\left( |\tilde{v}_{ij}(\eta)|^2 \right) = 4 N_\text{freq}^2 \sigma_\text{vis}^4.
\end{equation}
From Equation \ref{eq:power_spectrum_est}, and assuming constant noise across all baselines, the estimated power spectrum has noise variance
\begin{equation}
    \operatorname{Var}_\text{therm} \left[\hat{P}(k)\right] = \frac{4 N_\text{freq}^2 \sigma_\text{vis}^4 \sum_{ij\eta \in k} M^2_{ij}(\eta)}{\left[\sum_{ij\eta \in k} M_{ij}(\eta) \right]^2}.
\end{equation}
$M_{ij}(\eta)$ can take only values $0$ and $1$, so this is equivalent to
\begin{equation}
    \operatorname{Var}_\text{therm} \left[\hat{P}(k)\right] = \frac{4 N_\text{freq}^2 \sigma_\text{vis}^4}{N_\text{vis}(k)},
\end{equation}
where the denominator $N_\text{vis}(k) = \sum_{ij\eta \in k} M_{ij}(\eta)$ is simply the number of measurements that contribute to bin $k$. Figure \ref{fig:2d_samples} plots the number of samples as a function of $k_\perp$ and $k_\parallel$.

Here we have omitted any treatment of the visibilities' polarization and assumed a per-polarization analysis. Because the DSA-2000 antennas are dual-polarization, and the 21 cm signal is unpolarized, the XX and YY visibilities can be averaged to produce an estimate of the Stokes I signal.\footnote{More accurately, the average of the XX and YY visibilities estimates the ``pseudo'' Stokes I signal, which is a good approximation to the true Stokes I signal in the limit that XX and YY have orthogonal polarization projections across the field of view and near-identical beam responses. Delay spectrum analyses do not support true Stokes signal reconstruction. For a detailed discussion of polarized image reconstruction, see \citealt{Byrne2022a}.} Assuming independent thermal noise realizations from the two polarizations, the result is a factor of 2 improvement in the variance of the combined visibilities, producing a factor of 4 improvement in the thermal variance of the power spectrum. We therefore define the thermal variance as
\begin{equation}
    \operatorname{Var}_\text{therm} \left[\hat{P}(k)\right] = \frac{N_\text{freq}^2 \sigma_\text{vis}^4}{N_\text{vis}(k)}.
\end{equation}

\begin{figure}
    \centering
    \includegraphics[width=0.5\columnwidth]{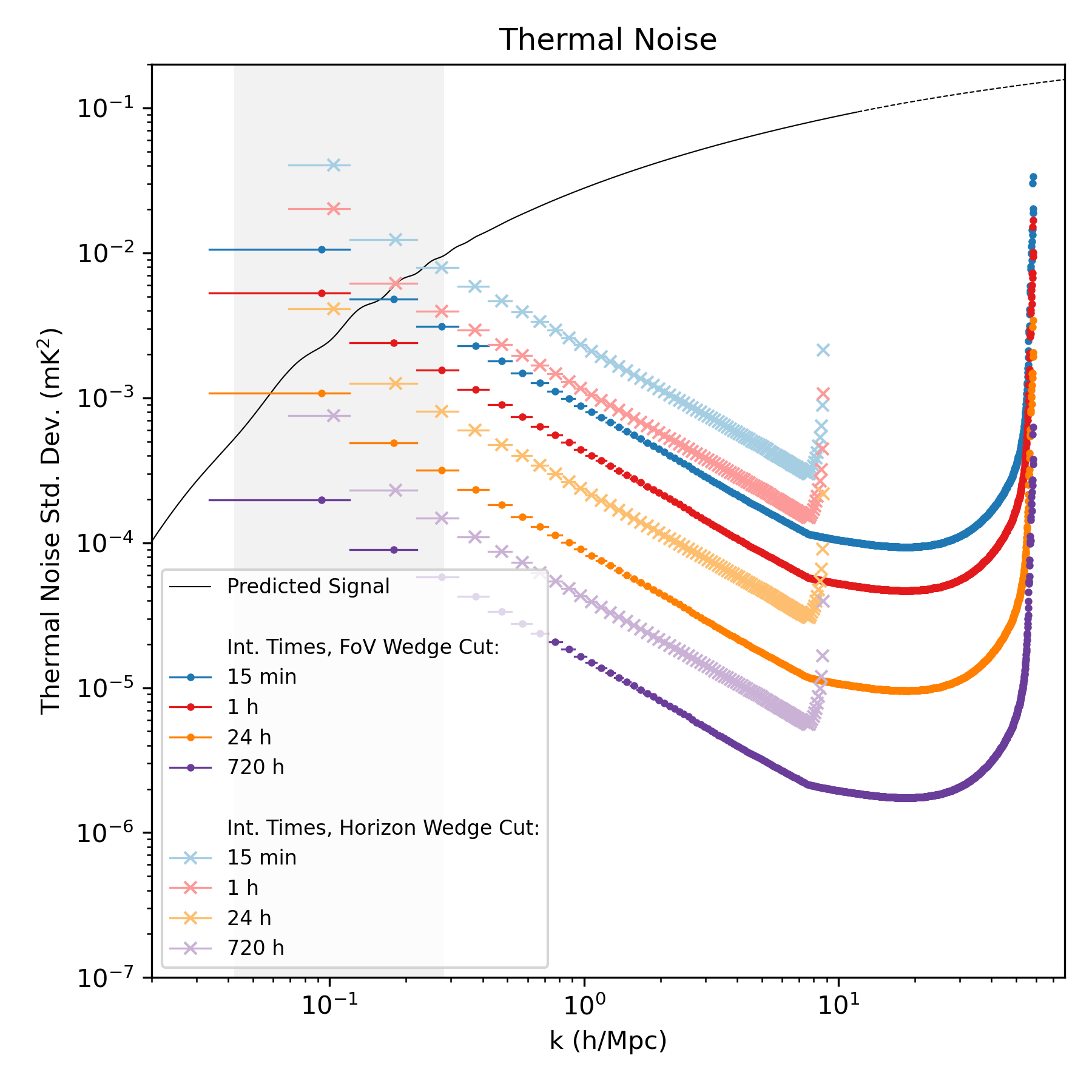}
    \caption{Expected thermal noise for power spectrum measurements made with the DSA-2000. We plot the thermal noise for bins of width $\Delta k = 0.1 h/\text{Mpc}$ at $0 \le z \le 1$. The different colors correspond to different integration times, from 15 minutes to a full season observation of 720 hours. We use the antenna configuration plotted in Figure \ref{fig:antlocs} and omit the effects of synthesis rotation, assuming zenith-pointed observations. We plot the results from two foreground avoidance strategies: circular markers exclude foreground wedge modes out to the limit of the field of view, and ``X'' markers use a more conservative foreground avoidance strategies that excludes all foreground wedge modes out to the horizon. The horizontal lines span the extent of the measurements that contribute to the given bin, and the marker is plotted at the $k$ location corresponding to the mean of those measurements. The shaded grey region indicates the modes corresponding to the first three Baryon Acoustic Oscillation (BAO) wiggles \citep{Bull2015}. The black line indicates the predicted power spectrum $P_\text{theory}(k, z=0.5)$ (see Equations \ref{eq:theory_ps} and \ref{eq:theory}). The solid black line is based on calculations from CAMB, and the dashed black line indicates interpolated values.}
    \label{fig:therm_noise}
\end{figure}

Figure \ref{fig:therm_noise} plots the predicted thermal noise levels for different integration times. We use linearly-spaced $k$ bins of width $\Delta k = 0.1 h/Mpc$ and plot results from two foreground avoidance strategies: excluding all foreground power within the field of view (up to the dashed white line in Figure \ref{fig:2d_samples}, denoted with circular markers in Figure \ref{fig:therm_noise}) and excluding modes up to the horizon threshold (the solid white line in Figure \ref{fig:2d_samples}, denoted with ``X'' markers in Figure \ref{fig:therm_noise}).

\subsection{Sample Variance}
\label{s:sample_variance}

The sample variance refers to the variance in the measurement due to the finite volume of space measured.\footnote{Note that while the terms are often used interchangeably, the ``sample variance'' is distinct from the ``cosmic variance,'' the theoretical minimium sample variance achievable if we were to measure the entire visible universe.} We measure finite regions on the sky across a limited redshift range, and in this section we estimate the resulting sample variance.

The first step in estimating the sample variance is to calculate the expected sample variance at each point in 3D power spectrum space, denoted $\boldsymbol{k}$. We define a signal $S(\boldsymbol{k})$ drawn from a complex circular Gaussian distribution with mean zero, and we define the power spectrum as $P(\boldsymbol{k}) = |S(\boldsymbol{k})|^2$. If the signal variance is
\begin{equation}
    \sigma^2 = \operatorname{Var}_\text{samp}\left( \Re\left[S(\boldsymbol{k})\right] \right) = \operatorname{Var}_\text{samp}\left( \Im\left[S(\boldsymbol{k})\right] \right),
\end{equation}
then
\begin{equation}
    \langle P(\boldsymbol{k}) \rangle = \langle |S(\boldsymbol{k})|^2 \rangle = \sigma^2,
\end{equation}
where the brackets $\langle \rangle$ denote the expectation value. The variance of the power spectrum is
\begin{equation}
    \operatorname{Var}_\text{samp}\left[ P(\boldsymbol{k}) \right] = \langle P^2(\boldsymbol{k}) \rangle - \langle P(\boldsymbol{k}) \rangle^2 = \sigma^4,
\end{equation}
from which it follows that
\begin{equation}
    \operatorname{Var}_\text{samp}\left[ P(\boldsymbol{k}) \right] = \langle P(\boldsymbol{k}) \rangle^2.
\end{equation}
We now need to convert this estimate of the point-by-point sample variance of the power spectrum signal into the sample variance of the binned power spectrum measured by our instrument. This requires calculating the characteristic correlation length of the measured signal and the volume of power spectrum space measured.

The correlation length across delay is given by
\begin{equation}
    \Delta \eta = \frac{1}{f_\text{max}-f_\text{min}}.
\end{equation}
For the $f_\text{min}$ and $f_\text{max}$ values given in Table \ref{tab:instrument_parameters}, this corresponds to $\Delta \eta = 1.37 \times 10^{-9} \, \text{s}$. Using the coordinate transformation given by Appendix \ref{app:cosmological_units} and the cosmological parameters from Table \ref{tab:cosmological_parameters}, this equates to a correlation length in $k_\parallel$ of $2.74 \times 10^{-3} \, h/\text{Mpc}$.

To estimate the \textit{uv} plane correlation length for a single snapshot image, we use the instrument's expected field of view of 30.0 deg$^2$, corresponding to a field of view diameter $\Delta \Theta = 6.18^\circ$. For each individual observation, the correlation length in the \textit{uv} plane is
\begin{equation}
    \Delta u = \frac{90^\circ}{\Delta \Theta},
\end{equation}
equivalent to half-wavelength spacing for a horizon-to-horizon observation or approximately 14.6 wavelengths for the DSA-2000's field of view. Transforming into cosmological units (see Appendix \ref{app:cosmological_units}), this becomes a $k_\perp$ correlation length of $1.53 \times 10^{-2} \, h/\text{Mpc}$.

We can define a correlation volume in 3D power spectrum space as
\begin{equation}
\Delta V = (\Delta u)^2 \Delta \eta.
\label{eq:corr_volume}
\end{equation}
In cosmological units, we estimate that the power spectrum signal is correlated across volumes of $6.43 \times 10^{-7} \, h^3/\text{Mpc}^3$.

Next, we need to estimate the total volume of 3D power spectrum space measured by our experiment, $V(k)$. Naively, this is equivalent to the volume of a spherical shell, but the calculation is complicated by the finite extent of the delay modes calculated and the foreground wedge, which mean that some power spectrum modes are excluded from the measurement. See Appendix \ref{app:sampling_volume} for the calculation of $V(k)$. Note that here we assume full sampling of the \textit{uv} plane. Our sample variance calculation does not consider the effect of ``holes'' in the \textit{uv} plane, where incomplete sampling would mean that the instrument measures a smaller effective volume of the 3D power spectrum space. However, this is a good approximation for the DSA-2000, which has near-complete \textit{uv} coverage.

For a given $k$ bin, the number of independent samples that contribute is given by
\begin{equation}
    N_\text{samp} = \frac{V(k) N_\text{fields}}{\Delta V},
\end{equation}
where $N_\text{fields}$ is the number of independent fields measured. The sample variance is then
\begin{equation}
    \operatorname{Var}_{\text{samp}}\left[ \hat{P}(k) \right] = \frac{\Delta V}{V(k) N_\text{fields}} P^2_\text{theory}(k, z),
\end{equation}
where $P_\text{theory}(k, z)$ is the predicted power spectrum from Equations \ref{eq:theory_ps} and \ref{eq:theory}.

\begin{figure}
    \centering
    \includegraphics[width=0.5\columnwidth]{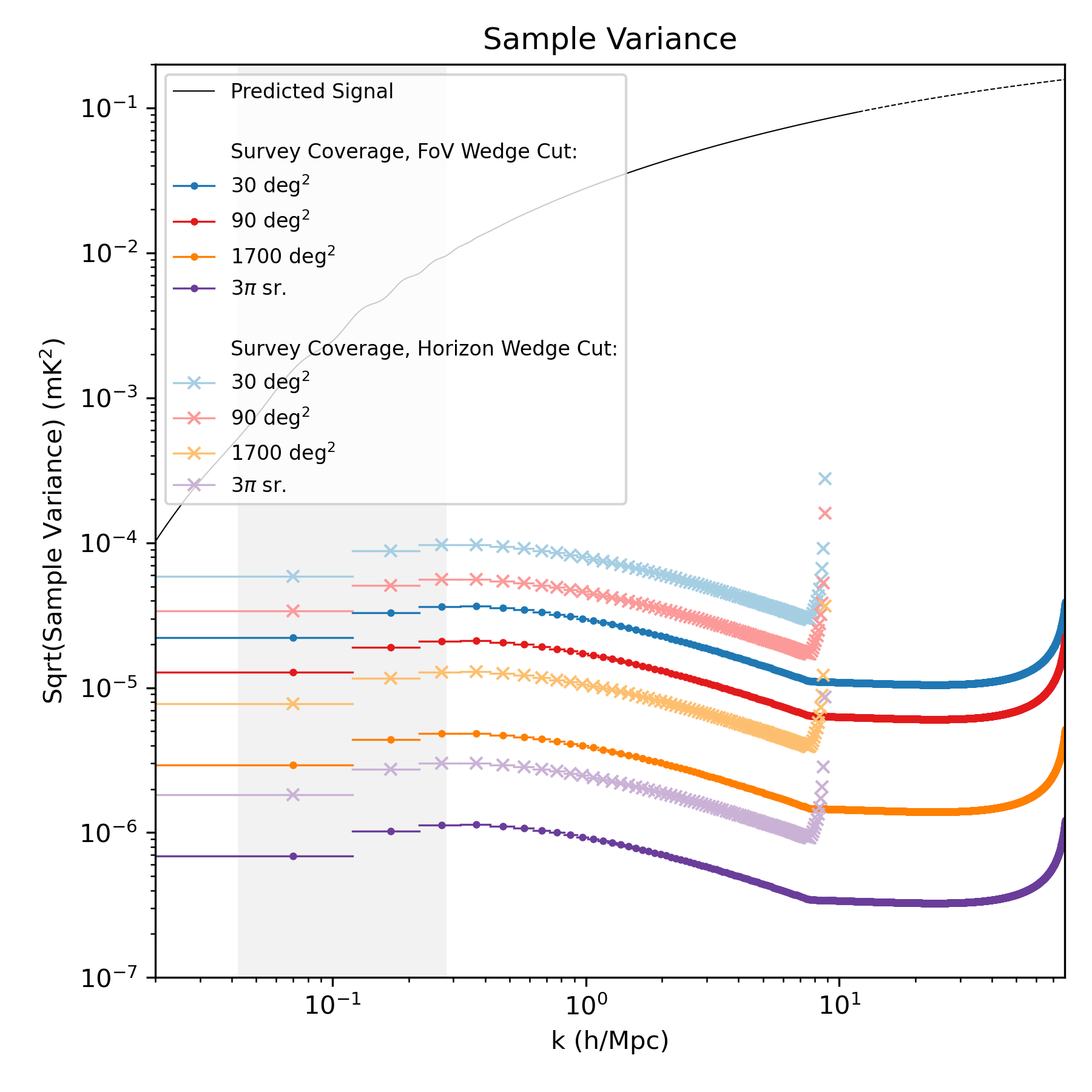}
    \caption{Expected sample variance for power spectrum measurements with the DSA-2000. Here we plot the square root of the sample variance, with units of mK$^2$. As in Figure \ref{fig:therm_noise}, we use bins of width $\Delta k = 0.1 h/\text{Mpc}$ at $0 \le z \le 1$. The horizontal lines span the extent of the bins, and the point is plotted its center. The black line indicates the predicted power spectrum $P_\text{theory}(k, z=0.5)$. The different colors correspond to different total fields of view. 30 deg$^2$ corresponds to the field of view for a single snapshot image, while $3 \pi$ steradians corresponds to the full sky visible from the DSA-2000's Northern Hemisphere location. Circles exclude foreground wedge modes out to the limit of the field of view; ``X''s use a more conservative foreground avoidance strategies that excludes all foreground wedge modes out to the horizon. The shaded grey region indicates the modes corresponding to the first three BAO wiggles. We find that the sample variance is a sub-dominant effect compared to the thermal noise.}
    \label{fig:sample_stddev}
\end{figure}

Figure \ref{fig:sample_stddev} plots the sample variance for field of view foreground avoidance (circles) and the more conservative horizon foreground avoidance (``X''s). We find that the sample variance is reduced for large survey areas, and that the thermal noise, plotted in Figure \ref{fig:therm_noise}, is the dominant source of measurement noise.

\subsection{Shot Noise}

An additional source of noise in the measured signal is shot noise, which emerges from the discrete nature of the galaxy halos. The density of galaxies in each volume element follows Poisson statistics, and the resulting noise contribution to the 21 cm signal depends on the galaxies' HI masses. 

A scale-invariant estimate of the shot noise is given by
\begin{equation}
    P_\text{shot}(z) = V_\text{box} \frac{\langle \sum_i (M_\text{HI}^i)^2 \rangle}{ \langle \sum_i (M_\text{HI}^i) \rangle^2},
\end{equation}
where $V_\text{box}$ is the sampled volume, $i$ indexes all HI sources within that volume, and $M_\text{HI}^i$ denotes the sources' HI mass \citep{Spinelli2020, Chen2021}. Evaluating this expression requires simulations of galaxy masses and halo occupation distributions, and the precise value depends on details of the simulation. Results from the Millenium II simulation, described in \citealt{Boylan-Kochin2009} and \citealt{Spinelli2020}, give $P_\text{shot}(z=0) = 46 \, \text{Mpc}^3/h^3$ and $P_\text{shot}(z=1) = 61 \, \text{Mpc}^3/h^3$; we infer $P_\text{shot}(z=0.5) \approx 53.5 \, \text{Mpc}^3/h^3$

In units of $\text{mK}^4$, the shot noise contribution to the signal variance in 3D power spectrum space is
\begin{equation}
\text{Var}_\text{shot}\left[ P(\boldsymbol{k}) \right] = \left( \frac{k^3}{2 \pi^2} [T_{21}(z)]^2 P_\text{shot}(z) \right)^2,
\end{equation}
where $T_{21}(z)$, the 21 cm brightness temperature, is given by Equation \ref{eq:brightness_temp}. As with the sample variance, averaging independent samples reduces the shot noise. We then get that the variance on the spherically averaged 1D power spectrum is
\begin{equation}
    \text{Var}_\text{shot} \left[ \hat{P}(k) \right] = \frac{\Delta V}{V(k) N_\text{fields}} \left( \frac{k^3}{2 \pi^2} [T_{21}(z)]^2 P_\text{shot}(z) \right)^2,
\end{equation}
where, as in \S\ref{s:sample_variance}, $\Delta V$ is the correlation volume (Equation \ref{eq:corr_volume}) and $V(k)$ is the total sampled volume in 3D power spectrum space (see Appendix \ref{app:sampling_volume}). $N_\text{fields}$ is the number of independent fields measured.

\begin{figure}
    \centering
    \includegraphics[width=0.5\columnwidth]{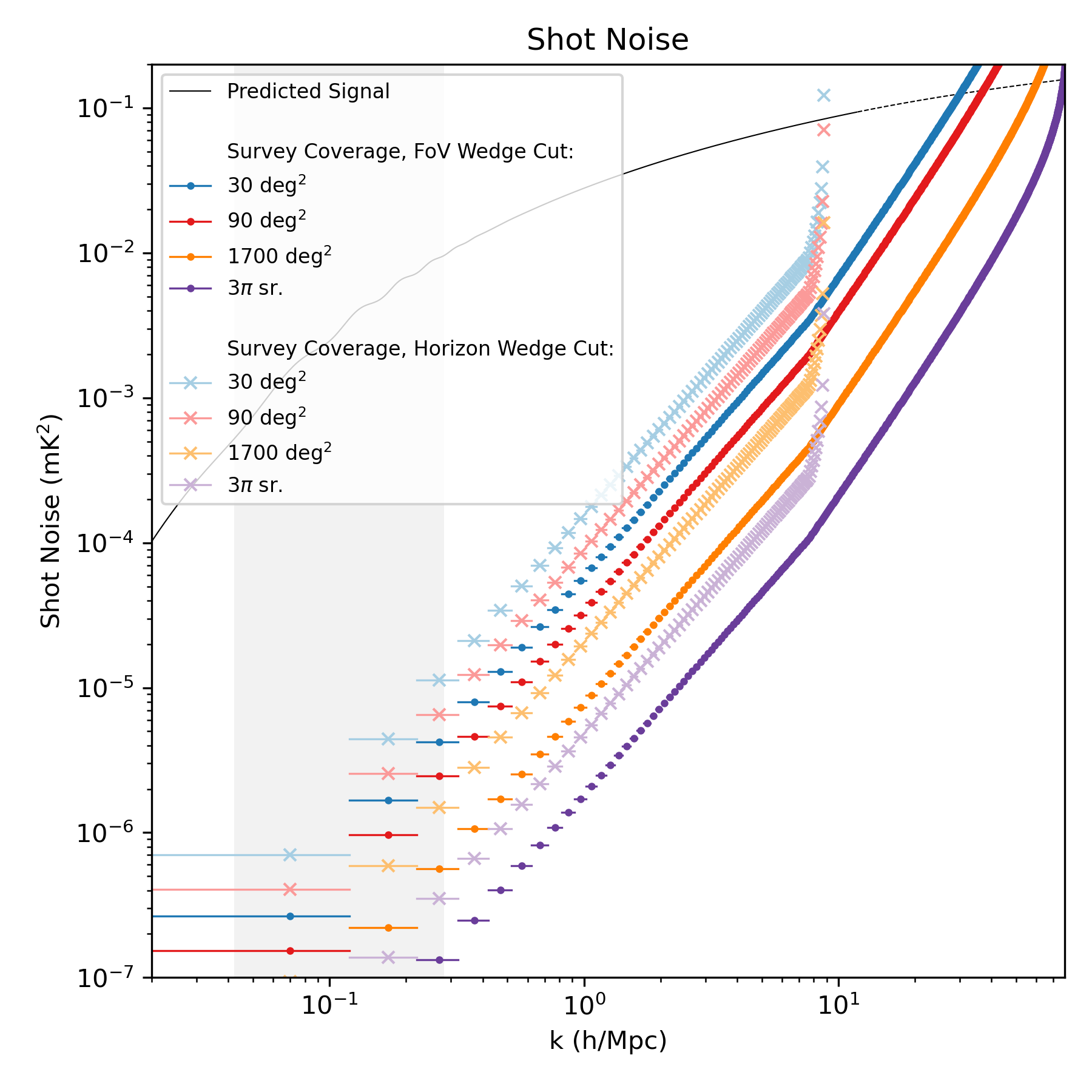}
    \caption{Expected shot noise for power spectrum measurements with the DSA-2000. As in Figures \ref{fig:therm_noise} and \ref{fig:sample_stddev}, we use bins of width $\Delta k = 0.1 h/\text{Mpc}$ at $0 \le z \le 1$. The horizontal lines span the extent of the bins, and the point is plotted its center. The black line indicates the predicted power spectrum $P_\text{theory}(k, z=0.5)$. The different colors correspond to different total fields of view for a more conservative foreground avoidance strategy (circles) and a field of view foreground avoidance strategy (``X''s). The shaded grey region indicates the modes corresponding to the first three BAO wiggles. Shot noise becomes a dominant effect at large $k$.}
    \label{fig:shot_noise}
\end{figure}

Figure \ref{fig:shot_noise} plots the shot noise for different total fields of view and two foreground avoidance strategies. We find that shot noise becomes a dominant source of noise for large $k$.

\subsection{Combined Sensitivity Estimate}

We define the combined signal variance as
\begin{equation}
    \operatorname{Var}\left[ \hat{P}(k) \right] = \operatorname{Var}_{\text{therm}}\left[ \hat{P}(k) \right] + \operatorname{Var}_{\text{samp}}\left[ \hat{P}(k) \right] + \operatorname{Var}_{\text{shot}}\left[ \hat{P}(k) \right].
\end{equation}
This quantifies the noise on the measurement but does not account for systematic error (see \S\ref{s:systematics}).

\begin{figure}
    \centering
    \includegraphics[width=0.9\columnwidth]{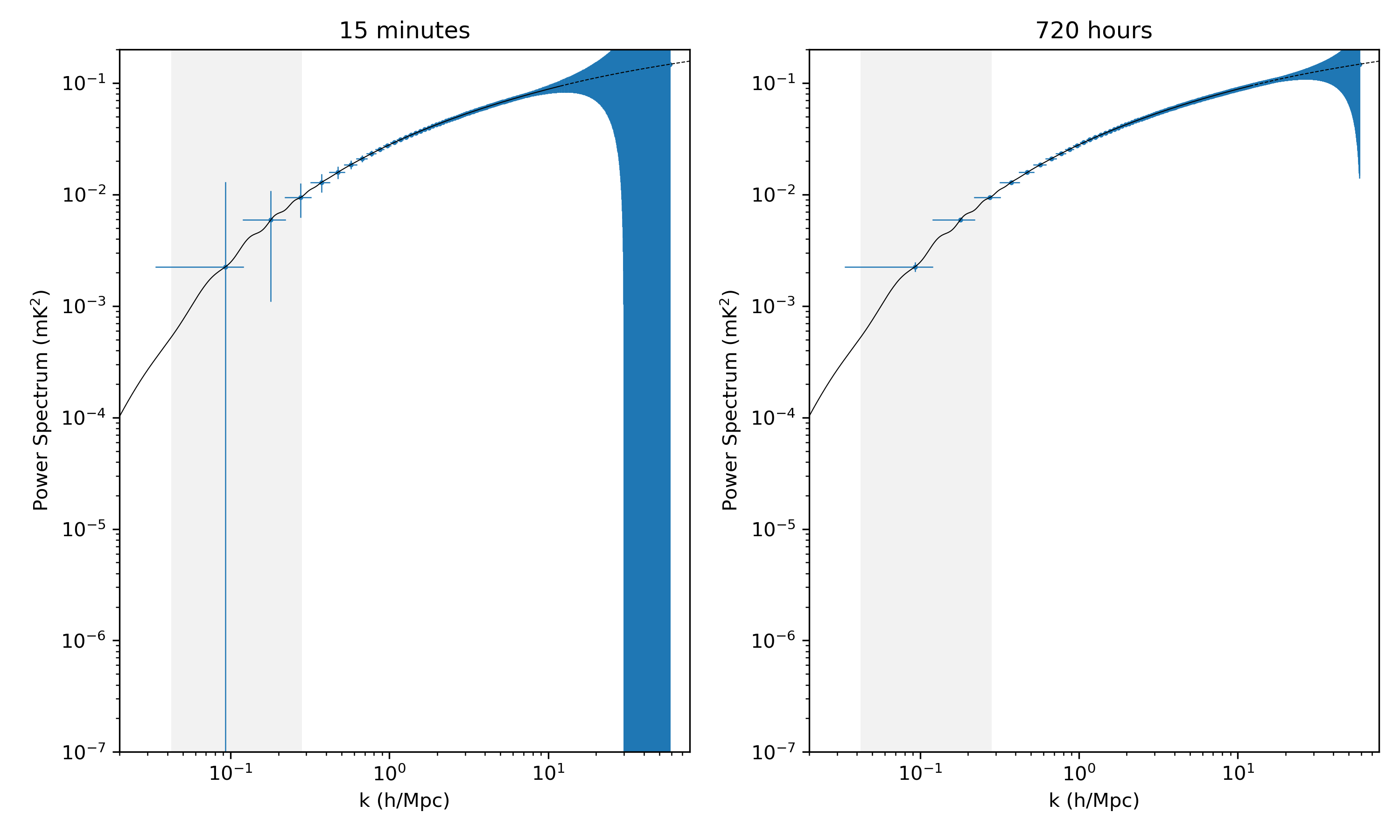}
    \caption{Expected power spectrum signal at z=0.5 with error bars overplotted. The left plot corresponds to a single 15 minute snapshot observation with a field of view of 30 deg$^2$. The right plot corresponds to a season of observing, with 720 hours of data total covering 1700 deg$^2$. Here we have excluded foreground wedge modes out to the limit of the field of view (dashed line in Figure \ref{fig:2d_samples}). The vertical blue error bars represent the predicted $1\sigma$ error, from combined thermal noise, sample variance, and shot noise effects. The horizontal blue lines span the extent of the measurements that contribute to the given bin, and the point is plotted at the $k$ location corresponding to the mean of those measurements. The shaded grey region indicates the modes corresponding to the first three BAO wiggles.}
    \label{fig:error_bars}
\end{figure}

Figure \ref{fig:error_bars} plots the combined variance as $1\sigma$ error bars, assuming that a field of view foreground wedge exclusion provides sufficient foreground mitigation. The left panel plots the error bars for a 15 minute snapshot observation, covering 30 deg$^2$ of the sky. The right panel includes 720 hours of data across 1700 deg$^2$, corresponding to a full season of observations at all local sidereal times (LSTs). We find that with just 15 minutes of data, the DSA-2000 has the sensitivity to produce a $5\sigma$ detection of the 21 cm power spectrum on modes $k=0.32-15.32 \, h/\text{Mpc}$ with $\Delta k = 0.1 \, h/\text{Mpc}$ resolution. With 720 hours of data, a $5\sigma$ detection is achievable on modes of $k=0.03-35.12 \, h/\text{Mpc}$.

\subsection{Baryon Acoustic Oscillations (BAOs)}
\label{s:bao}

\begin{figure}
    \centering
    \includegraphics[width=0.5\columnwidth]{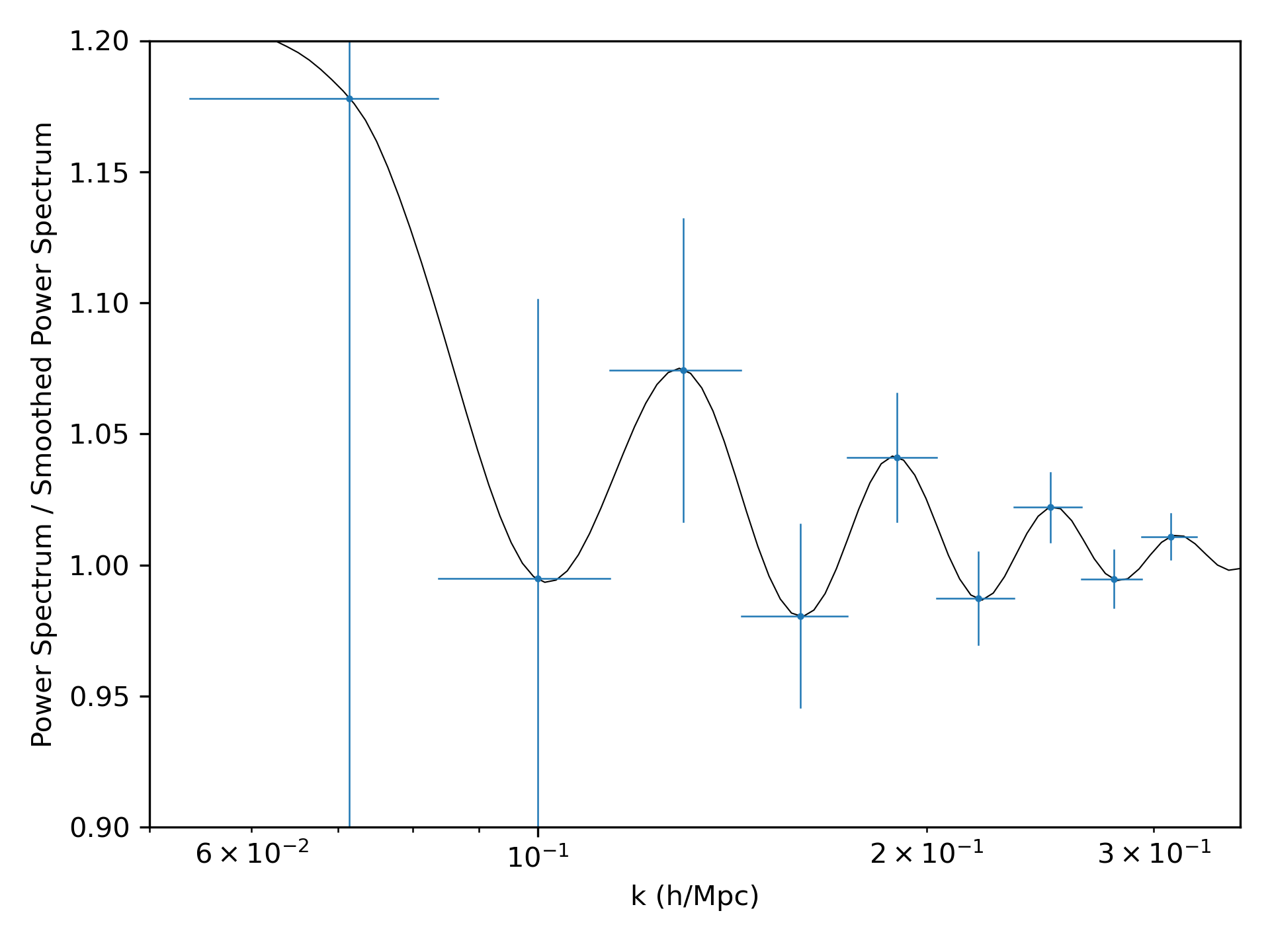}
    \caption{Expected BAO features with error bars overplotted for 720 hours of zenith-pointed data covering 1700 deg$^2$ of the sky. The vertical axis plots the predicted power spectrum $P_\text{theory}(k, z=0.5)$ divided by a smoothed function that has been convolved with a Gaussian. Here we have excluded foreground wedge modes out to the limit of the field of view. We use a resolution of $\Delta k = 0.03\, h/\text{Mpc}$ to resolve the BAO wiggles. The vertical blue error bars denote the predicted $1\sigma$ error, and the horizontal blue lines span the extent of the measurements that contribute to a given bin. We find that a season of data from the DSA-2000, with the nominal array configuration plotted in Figure \ref{fig:antlocs}, is insufficient to measure the BAO wiggles.}
    \label{fig:bao_error_bars}
\end{figure}

\begin{figure}
    \centering
    \includegraphics[width=0.5\columnwidth]{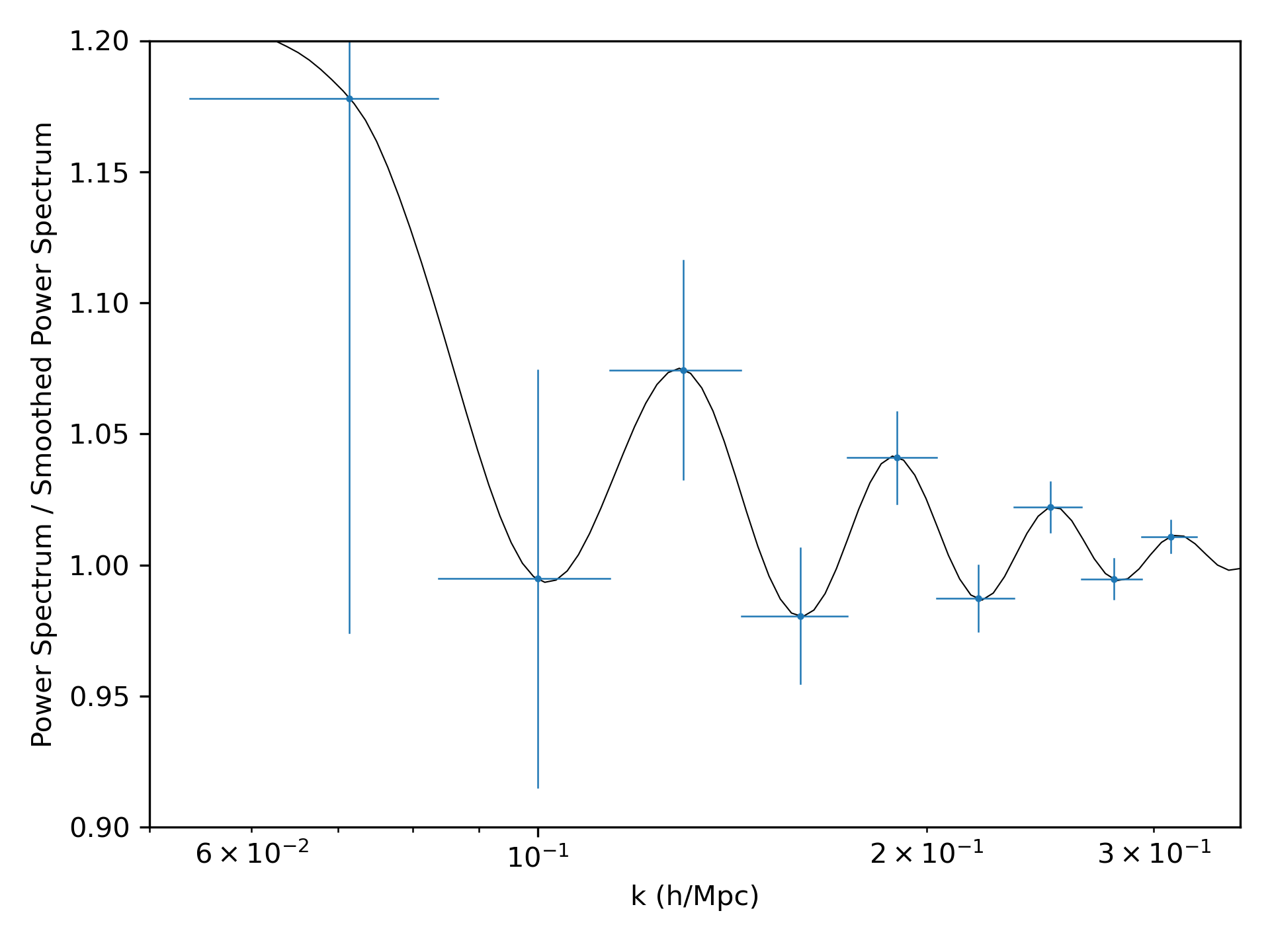}
    \caption{Expected BAO features with error bars overplotted. As in Figure \ref{fig:bao_error_bars}, we assume 720 hours of data covering 1700 deg$^2$, but here we assume the observations are taken at a zenith angle of $60^\circ$. The off-axis observations produce foreshortening of the baselines, allowing for better measurement of small $k$ modes and reduced error compared to the zenith pointed observations represented in Figure \ref{fig:bao_error_bars}. However, we still find that we would not have the sensitivity to measure the BAO wiggles.}
    \label{fig:bao_error_bars_offaxis}
\end{figure}

Of particular interest for low redshift 21 cm measurement are the BAO features, detectable as wiggles in the power spectrum signature. In Figures \ref{fig:bao_error_bars}, \ref{fig:bao_error_bars_offaxis}, and \ref{fig:bao_error_bars_core}, we have divided the predicted power spectrum signal by a smoothed signal to highlight the BAO features, plotted in black. The smoothed signal is simply the predicted power spectrum convolved with a Gaussian with FWHM of $0.06 \, h/\text{Mpc}$, equal to the approximate peak-to-peak extent of each BAO wiggle. To resolve these features, we require better resolution than the $\Delta k = 0.1 \, h/\text{Mpc}$ resolution used above. We use a resolution of $\Delta k = 0.03 \, h/\text{Mpc}$, equivalent to the peak-to-trough extent of the features, and Figure \ref{fig:bao_error_bars} plots the resulting $1\sigma$ error bars for the 720 hour and $1700 \, \text{deg}^2$ survey, assuming foreground wedge avoidance out to the field of view limit. We find that the nominal DSA-2000 array design, plotted in Figure \ref{fig:antlocs}, will not have the sensitivity to measure the BAO wiggles. This is because the DSA-2000 has a dearth of short baselines that sample the low $k$ modes of interest for BAO measurement.

Figure \ref{fig:bao_error_bars} assumes zenith-pointed observations. Off-axis observations can sample lower $k$ modes because of baseline foreshortening. In Figure \ref{fig:bao_error_bars_offaxis}, we plot error bars resulting from a simulation of observations taken at a zenith angle of $60^\circ$ toward the East. Although the off-axis pointing reduces the noise on these low $k$ modes, it is not sufficient to enable detection of the BAO wiggles.

\begin{figure}
    \centering
    \includegraphics[width=\columnwidth]{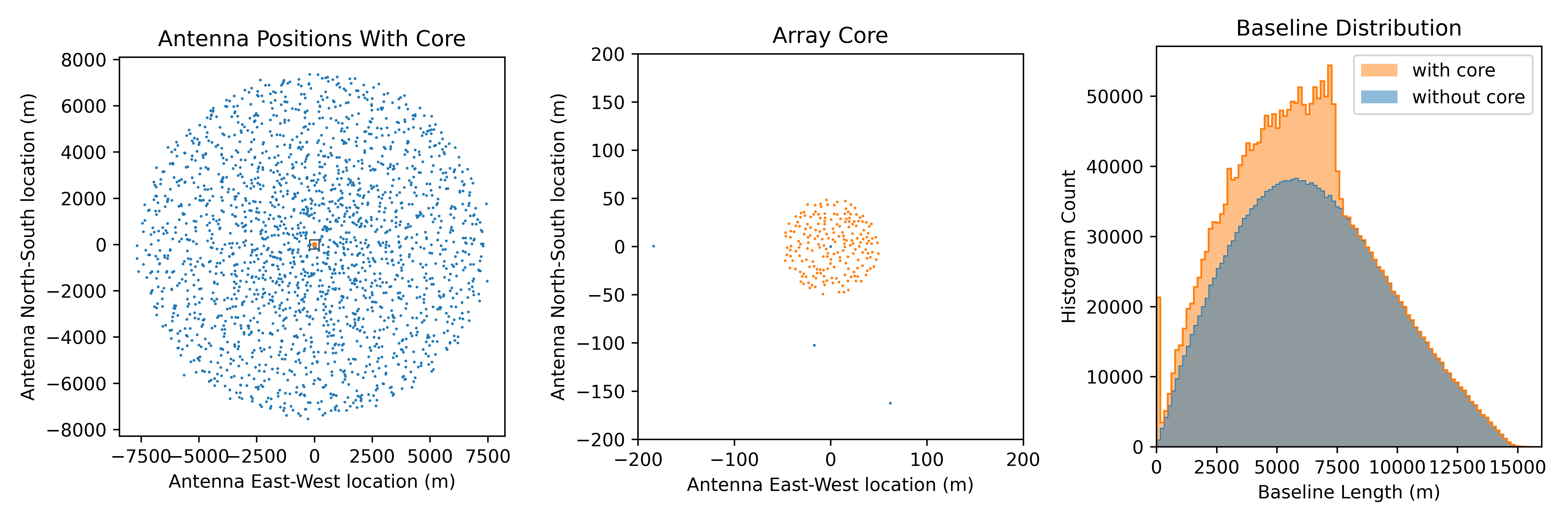}
    \caption{Supplementing the nominal array layout plotted in Figure \ref{fig:antlocs} with a dense core enhances the instrument's sensitivity to low $k$ modes. We explore a core of 200 additional antennas distributed randomly within a circular region of radius 50 m and with minimum antenna separation of 3 m. The resulting antenna layout is plotted in the left panel, where orange points represent the supplemented core antennas. The black outline in the left panel denotes the boundaries of the plot presented in the center panel. In the right panel we plot the resulting distribution of baseline lengths.}
    \label{fig:antlocs_core}
\end{figure}

\begin{figure}
    \centering
    \includegraphics[width=0.5\columnwidth]{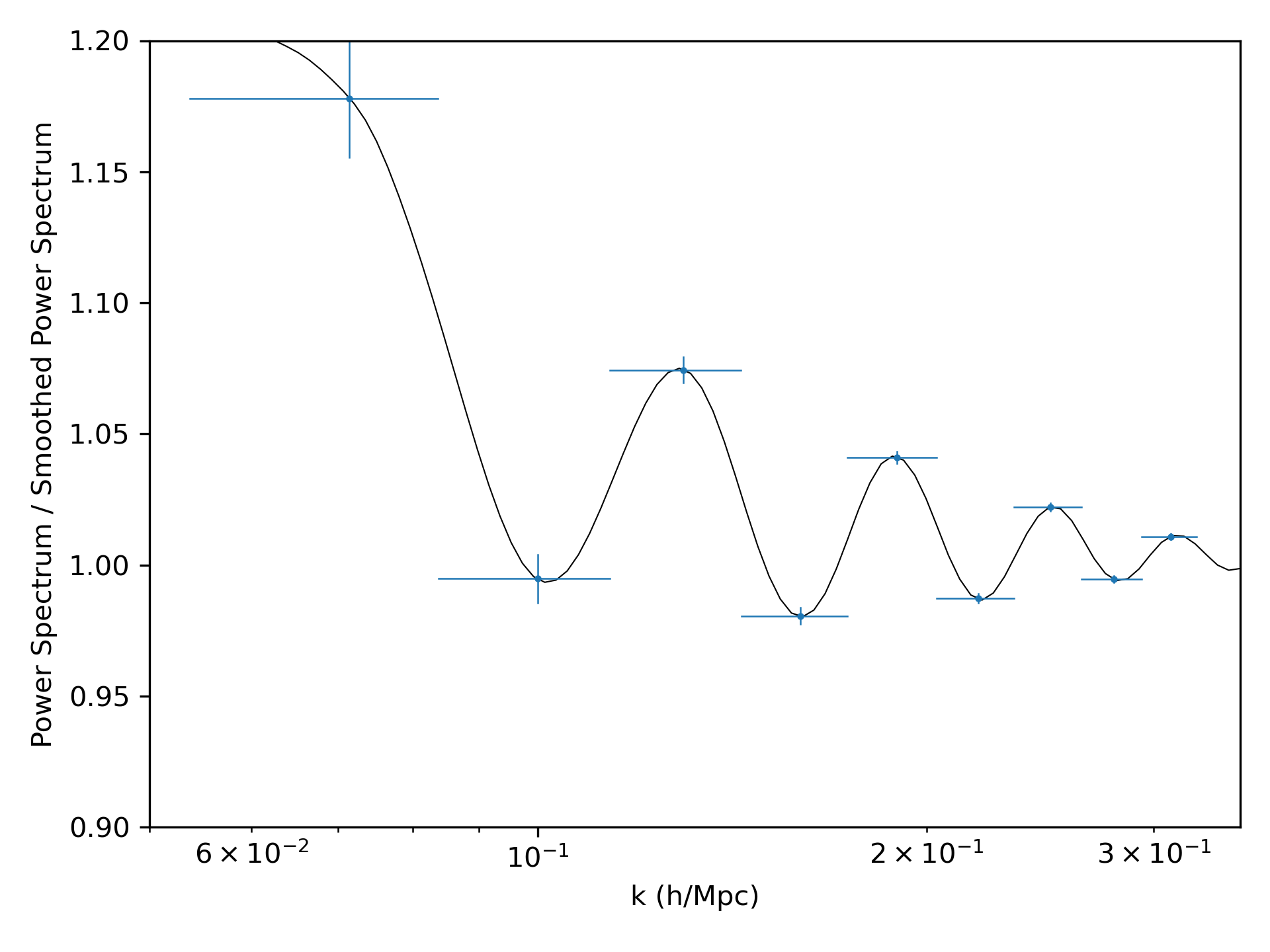}
    \caption{Supplementing the nominal DSA-2000 layout plotted in Figure \ref{fig:antlocs} with a dense core, plotted in Figure \ref{fig:antlocs_core}, would enable measurement of the BAO wiggles. As in Figures \ref{fig:bao_error_bars} and \ref{fig:bao_error_bars_offaxis}, we simulate 720 hours of data covering 1700 deg$^2$. As in Figure \ref{fig:bao_error_bars}, we assume zenith-pointed observations. We assume foreground avoidance out to the field of view limit of the foreground wedge.}
    \label{fig:bao_error_bars_core}
\end{figure}

Measurement of BAO wiggles would require supplementing the nominal DSA-2000 layout from Figure \ref{fig:antlocs} with a more densely-packed core sub-array. We explore the impact of a 200-antenna core on the sensitivity of measurement of the BAO wiggles. We simulate randomly placed antennas within a radius of 50 m with a minimum antenna separation of 3 m, plotted in Figure \ref{fig:antlocs_core}. The resulting $1\sigma$ error bars on the BAO wiggles are plotted in Figure \ref{fig:bao_error_bars_core}. The additional 200 close-packed core antennas enable the low-$k$ sensitivity needed to measure the BAO features.

\section{Systematics}
\label{s:systematics}

In \S\ref{s:sensitivity} we showed that the DSA-2000 has the necessary sensitivity to measure 21 cm power spectrum across a wide range of angular scales. However, sensitivity alone is not sufficient for enabling 21 cm intensity mapping, as systematics in the analysis can quickly overwhelm the faint signal. These systematics include calibration error, spectral mode-mixing, and contamination from RFI.

Calibration precision is a leading challenge for 21 cm intensity mapping experiments, as frequency-dependent gain errors as small as one part in $10^4-10^5$ can swamp the faint 21 cm signal with foreground contamination \citep{Barry2016}. The DSA-2000 will be uniquely calibratable, due to its many antennas and configuration optimized for imaging. An approximately 2000-antenna array produces over 2 million visibility measurements at each frequency and time, leading to an extremely constrained calibration problem. Calibratability is further enhanced by the DSA-2000's pseudo-random configuration. \citealt{Byrne2019} shows that non-regular arrays are more resilient to calibration error than regular arrays such as CHIME, CHORD \citep{Vanderlinde2019}, or HERA \citep{DeBoer2014}, where antennas are located on a repeating rectangular or hexagonal grid. This is because each visibility produced by a non-regular array captures an independent mode of the sky signal and, by extension, an independent mode of the sky model used in calibration. This holds true even when regular arrays are calibrated with redundant calibration (\citealt{Byrne2019}; for background on redundant calibration, see \citealt{Wieringa1992}, \citealt{Liu2010}, and \citealt{Dillon2020}). Furthermore, the DSA-2000's excellent imaging capabilities make it well-suited to self-calibration, where iterative calibration and imaging steps improve calibration precision \citep{Wieringa1992}. It could also benefit from unified calibration, a technique akin to self-calibration that is particularly effective for either regular arrays or non-regular arrays with dense \textit{uv} coverage, such as the DSA-2000 \citep{Byrne2021b}.

Precision calibration requires good knowledge of the direction-dependent beam response, as low-level beam errors introduce deleterious spectral calibration error \citep{Orosz2019, Joseph2020}. As a result, there has been significant investment in modeling the beam response of interferometers' antennas, and in extreme cases the gains must be calculated independently for different directions on the sky using sophisticated direction-dependent calibration algorithms \citep{Bhatnagar2008, Yatawatta2012, Yatawatta2015, Weeren2016, Patil2017, Kenyon2018}. The DSA-2000's uniform, fully steerable dishes are ringed with a screen, pictured in Figure \ref{fig:dish_image}. The screen lowers the system noise by mitigating ground spillover and reduces the antennas' mutual coupling, which can lead to non-uniform beam responses and spectral systematics \citep{Kern2019}. Since the dishes are fully steerable, the beam response can be directly measured by rastering across a bright source or employing beam holography techniques \citep{Berger2016}.

While the DSA-2000 is in principle exquisitely calibratable, achieving the optimal calibration precision is complicated by the fact that it will be infeasible to re-calibrate the raw data. The Radio Camera imaging approach reduces data in real time, enabling storage of the processed images rather than the raw visibilities. This is a departure from previous 21 cm intensity mapping experiments, which typically achieve calibration precision through iterative processing of the same visibility dataset. For the DSA-2000, the calibration pipeline must achieve the needed precision from the outset. To achieve the DSA-2000's full potential, the Radio Camera's calibration pipeline must incorporate state-of-the-art precision calibration techniques. Pipeline development should be informed by the strict calibration precision requirements for 21 cm intensity mapping.

Apart from its benefit to calibration, the DSA-2000's psuedo-random configuration, optimized for imaging, reduces systematics related to spectral mode-mixing. The DSA-2000 has an extremely well-behaved PSF with low sidelobe amplitudes (see Figure \ref{fig:psf}). This design was motivated by imaging performance requirements, as it enables accurate image-plane deconvolution \citep{Connor2022}. However, it also benefits spectral analyses such as 21 cm intensity mapping. PSF sidelobes are intrinsically frequency-dependent, and this gives rise to spectral leakage of the foregrounds into the foreground wedge, as discussed above in \S\ref{s:foreground_masking}. The amount of spectral leakage scales with the amplitude of the PSF sidelobes, meaning that the DSA-2000 would couple less foreground power into the foreground wedge than other 21 cm intensity mapping experiment. This could be further enhanced with foreground subtraction, leveraging the DSA-2000's high fidelity imaging capabilities. Further study is needed to explore the predicted amplitude of the DSA-2000's foreground wedge and determine if foreground wedge modes could be recovered, something that has thus far been infeasible in the field.

RFI contamination is a principal limitation for 21 cm analyses \citep{Barry2019b, Wilensky2020}. Even when the affected channels are flagged, RFI flags themselves produce spectral contamination of the signal \citep{Offringa2019, Wilensky2021, Chakraborty2022, Pagano2023}. The state of Nevada has among the lowest population densities in the country, and the DSA-2000 site selection has been driven by the relative RFI environment of the candidate sites. While intensive RFI studies have informed the site selection, further study is required to quantify the expected signal contamination level and its impact on the DSA-2000 science cases. While RFI will certainly pose a challenge for 21 cm intensity mapping, the DSA-2000 has the distinct advantage of enabling deep statistical RFI detection algorithms from its many antennas. This will enhance the capability of state-of-the-art RFI excision algorithms such as AOFlagger \citep{Offringa2015} and SSINS \citep{Wilensky2019}.

\section{Discussion}
\label{s:discussion}

The forthcoming DSA-2000 has a myriad of impactful science applications, and its many antennas, low system temperature, and fast survey speeds provide the sensitivity needed for 21 cm intensity mapping on a wide range of angular scales.

In \S\ref{s:sensitivity} we explore the effect of thermal noise, sample variance, and shot noise on observations made by the DSA-2000. We find that a season of observations, amounting to 720 hours of data, has the necessary sensitivity to produce a $5\sigma$ detection of the 21 cm power spectrum at $z \approx 0.5$ and scales of $k=0.03-35.12 \, h/\text{Mpc}$, with resolution of $\Delta k = 0.1 h/\text{Mpc}$. Furthermore, most of these modes are measurable with far less data. A 15 minute snapshot observation has the sensitivity to produce this detection at scales of $k=0.32-15.32 \, h/\text{Mpc}$.

The DSA-2000 is not the only instrument in the field with the necessary sensitivity for 21 cm intensity mapping. However, it features many more antennas than other near-redshift 21 cm intensity mapping experiments, including CHIME, CHORD, MeerKAT, and HIRAX. The sheer number of antennas, along with its pseudo-random array configuration and fully steerable dishes, make the DSA-2000 uniquely resilient to the systematic effects that limit other 21 cm intensity mapping experiments. Namely, the DSA-2000 has the potential to be calibrated more precisely than other instruments, reducing the spectral calibration error that is a dominant systematic for many 21 cm analyses.

Measurement of the 21 cm power spectrum with the DSA-2000 will lead to better constraints on cosmological and astrophysical parameters, which we forecast in a detailed Fisher matrix analysis in Mahesh et al.\ (in prep.). In general terms, the signal most directly traces the HI density fraction, $\Omega_\text{HI} b$ \citep{Bull2015, Santos2016, Pourtsidou2017, Vanderlinde2019, Liu2020}. It can also constrain the Hubble constant $H_0$, the dark energy density $\Omega_\Lambda$, the scalar spectral index $n_s$, the matter fluctuation parameter $\sigma_8$, neutrino mass, and the angular diameter distance  \citep{Bull2015, Villaescusa-Navarro2015, Pourtsidou2017, Padmanabhan2019}. When combined with other tracers, such as optical galaxy surveys, 21 cm intensity mapping overcomes the surveys' bias toward high-mass galaxies and captures faint, large-scale structure in galaxy filaments. This produces better constraint on astrophysical parameters including the HI halo mass slope $\beta$, the HI mass function cutoff, the scale-dependent optical galaxy bias, and the velocity dispersion $\sigma_v$ \citep{Padmanabhan2020, Paul2023}, with implications for understanding the baryon cycle and gas accretion \citep{Sun2019, Walter2020}. This makes 21 cm intensity mapping with the DSA-2000 highly complementary to upcoming galaxy surveys with DESI, Euclid, and the Rubin Observatory. Intensity mapping measurements will also synergize with the DSA-2000's own surveys of resolved HI sources.

Measurements of the BAO wiggles is a particularly compelling goal for near redshift 21 cm intensity mapping, as they would provide strong constraints on cosmic expansion history and $H_0$ \citep{CHIME2022}. The DSA-2000's nominal design (Figure \ref{fig:antlocs}) is not optimized for measuring low $k$, where we expect to find strong BAO signatures. As we show in \S\ref{s:bao}, supplementing this design with a dense core, such as the 200-antenna random core plotted in Figure \ref{fig:antlocs_core}, would supply additional short baselines and enable measurement of the BAO wiggles, as plotted in Figure \ref{fig:bao_error_bars_core}. The feasibility of constructing such a core remains to be seen.

The emergence of many-element radio arrays, enabled by advances in radio interferometric hardware and computing systems, ushers in a new era for 21 cm intensity mapping. Arrays such as the DSA-2000, with pseudo-random array configurations and well-behaved PSFs, are particularly suited for these measurements. In the coming years, the DSA-2000 could become a valuable addition to the field of near redshift 21 cm intensity mapping.

\section*{Acknowledgements}

This research received support through the generosity of Eric and Wendy Schmidt by recommendation of the Schmidt Sciences program. It is based upon work supported by the National Science Foundation under Award No.\ 2303952. Part of this research was carried out at the Jet Propulsion Laboratory, California Institute of Technology, under a contract with the National Aeronautics and Space Administration. Thank you to Jonathan Pober, Ari Cukierman, Yuping Huang, and the JPL Radio Cosmology Journal Club for their valuable input to this paper. 

\appendix

\section{Instrumental to Cosmological Unit Conversion}
\label{app:cosmological_units}

Throughout this paper we represent the mapping between instrumental and cosmological variables with the parameters $C_\parallel$ and $C_\perp$, defined as
\begin{equation}
    k_\parallel = C_\parallel \eta,
\end{equation}
where $k_\parallel$ are the line of sight power spectrum modes and $\eta$ is delay,
and
\begin{equation}
    \boldsymbol{k}_\perp = C_\perp \boldsymbol{u},
\end{equation}
where $\boldsymbol{k}_\perp = (k_x, k_y)$ are the power spectrum modes perpendicular to the line of sight and $\boldsymbol{u}$ are coordinates in the \textit{uv} plane. In this appendix we define $C_\parallel$ and $C_\perp$.

We transform between $\eta$ and $k_\parallel$ via the relationship 
\begin{equation}
    k_\parallel \approx \frac{2 \pi f_{21} E(z)}{D_H (1+z)^2} \eta
\end{equation}
\citep{Hogg1999}. Here $f_{21}$ is the 21 cm frequency and $D_H$ is the Hubble distance, defined as $D_H=c/H_0$ where $H_0$ is the Hubble constant. The expression $E(z)$ is given by
\begin{equation}
E(z) = \sqrt{\Omega_m(1+z)^3+\Omega_k(1+z)^2+\Omega_{\Lambda}}
\end{equation}
\citep{Morales2004}. 

Perpendicular to the line-of-sight, we transform from the baseline vector to cosmological modes via the relationship
\begin{equation}
    \boldsymbol{k}_\perp = \frac{2 \pi}{D_M(z)} \boldsymbol{u}
\end{equation}
where $D_M(z)$ is the transverse comoving distance \citep{Morales2004}. $\boldsymbol{u}$ is the \textit{uv} coordinate of a given baseline, given by $\boldsymbol{u} = \boldsymbol{b}/\lambda$ where $\boldsymbol{b}$ is the vector separation of the antenna pair and $\lambda$ is the wavelength, which we define at the center of the bandpass.

From \citealt{Hogg1999}, the transverse comoving distance is given by
\begin{equation}
  D_M=\begin{cases}
    D_H \frac{1}{\sqrt{\Omega_k}} \text{sinh}(\sqrt{\Omega_k}D_C/D_H), & \text{for $\Omega_k>0$}\\
    D_C, & \text{for $\Omega_k=0$}\\
    D_H \frac{1}{\sqrt{\Omega_k}} \text{sin}(\sqrt{\Omega_k}D_C/D_H), & \text{for $\Omega_k<0$}
  \end{cases}.
\end{equation}
$D_C$, the comoving distance, is
\begin{equation}
D_C = D_H \int_0^z \frac{dz'}{E(z')}.
\end{equation}
This integral is not analytically tractable, so we evaluate it numerically with the \texttt{scipy.integrate} function.

\section{Sampling Volume}
\label{app:sampling_volume}

The sample variance calculation presented in \S\ref{s:sample_variance} requires an estimate of the total volume of 3D power spectrum space measured by the experiment.

The theoretical maximum volume that contributes to a given power spectrum bin is given by the volume of a spherical shell:
\begin{equation}
    V(k) = \int_{k-\Delta k/2}^{k + \Delta k/2} dr \, r^2 \int_0^{\pi} d\phi \int_{-1}^{1} d (\cos \theta) = 2 \pi k^2 \, \Delta k + \frac{\pi}{6} (\Delta k)^3.
\end{equation}
Here $\theta$ corresponds to the polar angle, $\phi$ corresponds to the azimuthal angle, and $r$ corresponds to the radius. We integrate the azimuthal angle from 0 to $\pi$ only to account for the fact that the \textit{uv} plane is Hermitian, so only half the plane contributes independent values. While this is the maximum volume that could contribute to each measurement, in practice we need to exclude regions of power spectrum space outside the range of delay modes measured, as established by the instrument's frequency resolution, and within the foreground wedge, which is excluded to mitigate foreground contamination (see \S\ref{s:foreground_masking}).

The delay range measured is given by $-\eta_\text{max} \le \eta \le \eta_\text{max}$, where $\eta_\text{max} = \frac{1}{2 \Delta f}$. Transforming into cosmological units, this means that measured values of $k_\parallel$ fulfill
\begin{equation}
    |k_\parallel| \le \frac{C_\parallel}{2 \Delta f}.
\end{equation}
Using the instrument's frequency resolution of $\Delta f = 162.5 \, \text{kHz}$, as specified in Table \ref{tab:instrument_parameters}, and calculating $C_\parallel$ from Appendix \ref{app:cosmological_units}, we find that we measure values $|k_\parallel| \le 6.01 \, h/\text{Mpc}$. Converting into spherical coordinates, where $k_\parallel = r \cos \theta$, we get that
\begin{equation}
    |\cos \theta| \le \frac{C_\parallel}{2 \Delta f \, r}.
\label{eq:delay_max_inequality}
\end{equation}

Next, we can account for the effects of foreground avoidance. In \S\ref{s:foreground_masking} we note that the only measurements that contribute to the power spectrum estimate fulfill
\begin{equation}
    |\eta| \gtrsim \sin \omega \, |\boldsymbol{b}_{ij}|/c,
\end{equation}
where $|\boldsymbol{b}_{ij}|$ is the baseline length and $\sin \omega$ determines the extent of the foreground wedge cut. This is equivalent to 
\begin{equation}
|\eta| \gtrsim \sin \omega \, \sqrt{u^2 + v^2}/f
\end{equation}
or
\begin{equation}
    |k_\parallel| \gtrsim \frac{\sin \omega \, C_\parallel}{C_\perp f} \, |k_\perp|,
\end{equation}
where $C_\parallel$ and $C_\perp$ are the unit conversion parameters defined in Appendix \ref{app:cosmological_units}. Letting $f=1.06 \, \text{GHz}$, the central frequency, the coefficient $\frac{C_\parallel}{C_\perp f} = 1.83$. We now convert this inequality into spherical coordinates, using the relationships $k_\parallel = r \cos \theta$ and $k_\perp = r \sin \theta$. We then require that
\begin{equation}
     |\cos \theta| \gtrsim \frac{\sin \omega \, C_\parallel}{C_\perp f} \sin \theta.
\end{equation}
Using the identity $\sin \theta = \sqrt{1-\cos^2 \theta}$ and solving for $|\cos \theta|$, we get that
\begin{equation}
     |\cos \theta| \gtrsim \frac{\sin \omega \, C_\parallel}{\sqrt{\cos^2 \omega \,C_\parallel^2 + C_\perp^2 f^2}}.
\end{equation}

We evaluate $V(k)$, the volume of 3D power spectrum space measured, by expressing these inequalities as integration limits. Because the integral is symmetric about $k_\parallel=0$ (or, equivalently, $\cos \theta=0$), we can evaluate the integral for positive $\cos \theta$ and account for the negative region with a factor of 2. When $k + \frac{\Delta k}{2} \le \frac{C_\parallel}{2 \Delta f}$ we need to only consider the foreground wedge cut, as the spherical shell lies fully within the range of delay values measured. We then get that
\begin{equation}
    V(k) = 2 \int_{k-\Delta k/2}^{k + \Delta k/2} dr \, r^2 \int_0^{\pi} d\phi \int_{\frac{\sin \omega \, C_\parallel}{\sqrt{\cos^2 \omega \,C_\parallel^2 + C_\perp^2 f^2}}}^{1} d (\cos \theta).
\end{equation}
When $k + \frac{\Delta k}{2} > \frac{C_\parallel}{2 \Delta f}$ we need to account for both the foreground wedge and the delay range limit. For $k - \frac{\Delta k}{2} \ge \frac{C_\parallel}{2 \Delta f}$, the integral becomes
\begin{equation}
    V(k) = 2 \int_{k-\Delta k/2}^{k + \Delta k/2} dr \, r^2 \int_0^{\pi} d\phi \int_{\frac{\sin \omega \, C_\parallel}{\sqrt{\cos^2 \omega \,C_\parallel^2 + C_\perp^2 f^2}}}^{\frac{C_\parallel}{2 \Delta f \, r}} d (\cos \theta),
\end{equation}
and when $k - \frac{\Delta k}{2} < \frac{C_\parallel}{2 \Delta f} < k + \frac{\Delta k}{2}$ we evaluate
\begin{equation}
    V(k) = 2 \int_{k-\Delta k/2}^{\frac{C_\parallel}{2 \Delta f}} dr \, r^2 \int_0^{\pi} d\phi \int_{\frac{\sin \omega \, C_\parallel}{\sqrt{\cos^2 \omega \,C_\parallel^2 + C_\perp^2 f^2}}}^{1} d (\cos \theta) 
    + 2 \int_{\frac{C_\parallel}{2 \Delta f}}^{k + \Delta k/2} dr \, r^2 \int_0^{\pi} d\phi \int_{\frac{\sin \omega \, C_\parallel}{\sqrt{\cos^2 \omega \,C_\parallel^2 + C_\perp^2 f^2}}}^{\frac{C_\parallel}{2 \Delta f \, r}} d (\cos \theta).
\end{equation}
Evaluating these integrals with \textit{Mathematica} software, we get that
\begin{equation}
V(k) = \begin{cases}
    \left[ 2 \pi k^2 \, \Delta k + \frac{\pi}{6} (\Delta k)^3 \right] \left(1 - \frac{\sin \omega \, C_\parallel}{\sqrt{\cos^2 \omega \,C_\parallel^2 + C_\perp^2 f^2}} \right), & k + \frac{\Delta k}{2} \le \frac{C_\parallel}{2 \Delta f} \\
    \pi k \, \Delta k \frac{C_\parallel}{\Delta f} - \left[ 2 \pi k^2 \, \Delta k + \frac{\pi}{6} (\Delta k)^3 \right] \frac{\sin \omega \, C_\parallel}{\sqrt{\cos^2 \omega \,C_\parallel^2 + C_\perp^2 f^2}}, & k - \frac{\Delta k}{2} \ge \frac{C_\parallel}{2 \Delta f} \\
    \frac{\pi}{8} \frac{C_\parallel}{\Delta f} \left( 2k + \Delta k \right)^2 - \frac{\sin \omega \, C_\parallel}{\sqrt{\cos^2 \omega \,C_\parallel^2 + C_\perp^2 f^2}} \left[ 2 \pi k^2 \Delta k + \frac{\pi}{6}(\Delta k)^3 \right] + \frac{\pi}{12} \left(\Delta k - 2 k \right)^3 - \frac{\pi}{24} \frac{C_\parallel^3}{(\Delta f)^3}, 
    & k - \frac{\Delta k}{2} < \frac{C_\parallel}{2 \Delta f} < k + \frac{\Delta k}{2} \\
\end{cases}.
\end{equation}


\bibliography{sample631}{}
\bibliographystyle{aasjournal}



\end{document}